\documentclass[a4paper,printer]{aa}
\usepackage{graphicx}
\usepackage[fleqn]{amsmath}
\usepackage{txfonts}
\usepackage{natbib}
\usepackage[latin1]{inputenc}
\usepackage[T1]{fontenc}
\usepackage{algorithm,algorithmic}
\usepackage{subfigure}
\usepackage{enumerate}

\def\INTEGRALSPI{\textit{INTEGRAL/SPI}}
\def\INTEGRAL{\textit{INTEGRAL}}
\def\IBIS{IBIS}
\def\SPI{SPI}
\def\INTEGRALIBIS{\textit{INTEGRAL/IBIS}}

\def\SWIFTBAT{Swift/BAT}

\def\timebins{`time bins'}

\def\imagredchi{\(\chi_r^{2(i)}\)}
\def\dataredchi{\(\chi_r^{2}\)}
\def\Nseries{L}
\def\Likeh{\mathcal{L}}

\authorrunning{Bouchet et al.}
\titlerunning{\INTEGRALSPI{} light curve through data segmentation}
\begin{document}

\title{\INTEGRALSPI{} 
\thanks{Based on observations with INTEGRAL, an ESA project with instruments and science data centre funded by ESA member states (especially the PI countries: Denmark,      France, Germany, Italy, Spain, and Switzerland), Czech Republic and Poland with participation of Russia and the USA.} data segmentation to retrieve source intensity variations
}
\author{L.~Bouchet\inst{1,2}, P. R.~Amestoy\inst{3}, A.~Buttari\inst{4}, F.-H.~ Rouet\inst{3,5} \and  M.~Chauvin\inst{1,2}}

\institute{\(^1\)Universit\'e de Toulouse, UPS-OMP, IRAP,  Toulouse, France \\
\email{lbouchet@irap.omp.eu}\\  
\(^2\)CNRS, IRAP, 9 Av. colonel Roche, BP 44346, F-31028 Toulouse cedex 4, France \\
\(^3\)Universit\'e de Toulouse, INPT-ENSEEIHT-IRIT, France \\
\(^4\)CNRS-IRIT, France\\
\(^5\)Lawrence Berkeley National Laboratory, Berkeley CA94720, USA}

\date{Received May 15, 2012; Accepted April 22, 2013}
 
\abstract
{The \INTEGRALSPI{}, X/\(\gamma\)-ray spectrometer (20 keV - 8 MeV)  is an instrument for which
recovering
source intensity variations is not straightforward and can constitute a  difficulty for data analysis. In most cases, determining the source intensity changes between exposures is largely based on a priori information.}
{We propose  techniques that help to overcome the difficulty related to source intensity variations, which make this step  more rational.
In addition, the constructed ``synthetic'' light curves should permit us to obtain a sky model that describes the data better and optimizes the source signal-to-noise ratios.}
{For this purpose, the time intensity variation of each source was modeled as a combination of piecewise segments of time 
during which a given source exhibits a constant intensity. To optimize the signal-to-noise ratios, the number of  segments was minimized.
We present a first method that takes advantage of previous time series that can be obtained from another instrument
on-board the \INTEGRAL{} observatory. A data segmentation algorithm was then used to synthesize the time series into segments. 
The second method no longer needs external light curves, but solely \SPI{} raw data.
For this, we developed a  specific algorithm that involves the \SPI{} transfer function.}
{The time segmentation algorithms that were developed solve a difficulty inherent to the \SPI{} instrument, 
which is the intensity variations of sources between exposures, and it allows us to obtain more information about the sources' behavior.}
{}


\maketitle

\section{Introduction}

The \SPI{} X/\(\gamma\)-ray spectrometer, on-board the \INTEGRAL{} observatory is dedicated to the analysis of both point-sources and diffuse emission \citep{Vedrenne03}.
The sky imaging is indirect and relies on a coded-mask aperture associated to a specific observation strategy that is based on a dithering procedure.
Dithering is needed since a single exposure does 
not always provide enough information or data to reconstruct the sky region viewed through the \(\sim\)\(30^{\circ}\) field-of-view, 
which may contain hundreds of sources for the most crowded regions.
It consists of small shifts in the pointing direction between exposures, the grouping of which allows us
to increase the amount of available information on a given sky target through a growing set of ``non-redundant'' data. 
The standard data analysis consists of adjusting a model of the sky convolved with the instrument transfer function plus instrumental background to the data.
\\
However, source intensities vary between exposures.
Thus, a reliable modeling of source variability, at least of the most intense ones, is needed 
to accurately model the data and intensity measurement of the sources.
In addition, ignoring these intensity variations can prevent the detection of the weakest sources.
\\
Ideally, the system of equations that connects the data to the sky model (unknowns) should be solved for intensities as well as for the variability timescales of 
sources and background. 
We propose methods based on data segmentation to fulfill these requirements as best as possible and we report in detail on 
the improved description and treatment of source variability during the data reduction process.
%
\section{Functioning of SPI: necessity of modeling the intensity variations of sources}
\label{sec:instrument}

\subsection{Instrument}
\SPI{} is a spectrometer provided with an imaging system sensitive to both point-sources and extended source/diffuse emission.  The instrumental characteristics and performance are described in \citet{Vedrenne03} and \citet{Roques03}. Data are collected between 20 keV and 8 MeV using 19 high-purity Ge detectors illuminated by the sky through a coded mask. The resulting field-of-view (FoV) is \(\sim\)\(30^{\circ}\).
The instrument can locate intense sources with an accuracy of a few arc minutes \citep{Dubath05}. \\
\subsection{\SPI{} imaging specifics and necessity of dithering}
\label{sec:dithering:mask}
A single exposure provides only 19 data values (one per Ge detector), which is not enough to correctly sample the sky image.
Indeed, dividing the sky into \(\sim\)\(2^{\circ}\) pixels (the instrumental angular resolution) yields about 225 unknowns for a \(30^{\circ}\) FoV (i.e., \( (30^{\circ}/2^{\circ})^2\)).
Hence, reconstructing the complete sky image enclosed in the FoV is not possible from a single exposure because the related system of equations is undetermined.
From these 19 data values, and the parameters describing the instrumental background we can, at best, determine the intensity of only a few
sources.
This is sufficient for sparsely populated sky regions and for obtaining a related coarse image (the intensities of the few pixels corresponding to the source positions).
However, some  of the observed sky regions are crowded (the Galactic center region, for example, contains hundreds of sources) and obtaining more data is mandatory for determining all parameters 
of the sky model.
The dithering technique (Sect.~\ref{sec:dithering:dithering}) allows to obtain this larger set of data through multiple exposures.
\subsection{Dithering}
\label{sec:dithering:dithering}
By introducing multiple exposures of the same field of the sky that are shifted by an offset that is small compared to the FoV size, 
the number of data related to this sky region is increased.
In the dithering strategy  used for \INTEGRAL, the spacecraft continuously follows a dithering pattern throughout an observation \citep{Jensen03}. In general, the exposure direction varies around the pointed target by steps of \(2^{\circ}\) within a five-by-five square or a seven-point hexagonal pattern. 
An exposure lasts between 30 minutes and 1 hour.\\
Through grouping of exposures related to a given sky region, 
the dithering technique allows one to increase the number of ``non-redundant'' 
data points\footnote{The measurements correspond to data points, that are not completely independent, if one considers that the shadow (of the mask projected onto the camera by a source) just moves as a function of the pointing direction of the exposure.\\
Dithering is also used to extend the FoV and to separate the background from the source emission. It does not always
provide a better coding of the sky.}
without significantly  increasing the size of the  FoV that is spanned by these  exposures.
This provides enough data to recover the source intensity by solving the related system of equations.
The source intensity variations within an exposure do not constitute a problem with the coded mask imaging system.
The difficulty comes from the intensity variations between exposures.
Hence, in our case, sources are variable on various timescales, ranging from an hour (roughly the duration of an exposure) to years.
This information should be included in the system of equations to be solved.
\subsection{Handling source intensity variation}
\label{sec:dithering:light curves}
We chose to model the intensity variation of a source as a succession of piecewise constant segments of time. In each of the segments (also called \timebins), the intensity of the source is supposed to be stable.
The higher energy bands (E $\ga$ 100 keV)
contain a few emitting/detectable sources. Their intensities are rather stable with time according to the statistics and 
only a few piecewise constant segments of time or \timebins\ are needed to model source intensity variations.\\
At lower energies (E $\la$ 100 keV), the source intensity varies more rapidly. 
When inappropriate timescales are used for the sources contained in the resulting large FoV (\( > 30^{\circ}\))
and for the background, the model does not fit the data satisfactorily (the chi-squares of the corresponding least-squares problem (Appendix~\ref{app:phystomath}) can be relatively high).
Moreover, allowing all sources to vary on the timescale of an exposure is an inappropriate strategy because, for crowded regions of the sky the problems to be solved are again undetermined in most cases.
Generally, to estimate the source variability timescale, a crude and straightforward technique consists of testing several mean timescale values until
the reduced chi-square of the associated least-squares problem is about 1 or does not decrease anymore. Unfortunately, this method is rapidly limited. \\
Including variability in the system of equations always increases the number of unknowns that need to be determined (Appendix~\ref{app:phystomath}) since the intensity in each segment of time (\timebins)  is to be determined. Using too many \timebins{} will increase the solution variance and does not necessarily produce the best solution. Similarly, it is better to limit the number of \timebins\ to the strict necessary minimum to set up a well-conditioned system of equations.\\
When one manually defines the \timebins{}, one might soon be overwhelmed by the many timescales to be tested and the number of sources. 
It is difficult to simultaneously search for the variability of all sources contained in the instrument FoV and to test the various combinations of 
\timebins{} (not necessarily the same length or duration, location in time).
As a consequence, constructing light curves turns out to be rather subjective and tedious, and relies most of the time on some a priori knowledge.
To make this step more rational, we propose methods based on a partition of the data into segments, to model source intensity variations between exposures.
%
\section{Material}
\label{sec:mathandmeth}
%
The objective is to find the \timebins\ or segment sizes and locations corresponding to some data. This is related to the partition of an interval of 
time into segments. We propose two methods.
The first one, called 'image space', relies on some already available light curves (or equivalently on a time series)
\footnote{In some cases it is also possible to use \SPI{} light curves directly, for example when the  
FoV  contains only a few sources. A first analysis exposure-by-exposure provides a light curve.}.
In our application the time series come mainly from 
\INTEGRALIBIS{} and from \SWIFTBAT{} \citep{Barthelmy05}.
The purpose is to simplify an original light curve to maximize the source signal-to-noise ratio (S/N), hence  to reduce the number of \timebins\ through  
minimizing of the number of time segments\footnote{The number of naturally defined \timebins\,
using for example the \IBIS\ `imager', is the number of exposures (hence, exposure time
scales), which can exceed the number of \SPI{} data  if we have more than 19 sources (which corresponds to the number of detectors) in the FoV varying on the exposure timescale), leading to an undetermined system of equations. In short, these light curves on the exposure timescale cannot be used directly.}.
These \timebins{} are used to set up the \SPI{} system of equations. This partitioning is made for all sources in the FoV.
This algorithm  is not completely controlled by \SPI{} data, but at least it allows one to obtain a reasonable
agreement (quantified by the chi-square value) between the data and the sky model.\\
The second method, called `data space',  starts from the \SPI{} data and uses the instrument transfer function. In contrast to the `image-space' method where the segmentation is made for a single source only,
the `data-space' method performs the time interval partitioning simultaneously for all sources in the FoV and is used to separate their individual contributions to the data. While this is more complex, 
the great advantage is that it is based solely on \SPI{} data (Sect.~\ref{sec:theory:dataspace}).
%
\begin{table}[!ht]
\caption{Index of mathematical symbols.}
\begin{tiny}
\renewcommand{\tabcolsep}{0.3em} 
\begin{tabular}{lp{0.389\textwidth}}
\(n_p\)            & number of exposures      \\
\(n_d\)            & number of detectors (data) per exposure\\
                   & a vector of length \(n_p\) \\
\(N\)              & number of parameters (to determine)  \\
\(\Nseries\)        & number of data points in a time series \\
\(M\)              & number of data points \(M= \sum_{i=1}^{n_p} n_d(i)\)\\
\(D_{p}\)          & data recorded during the exposure (pointing) \(p\) \\
                   & \(D_{p}\) is a vector of length \(n_d(p)\) \\
\(y\)              & data sorted in sequential order (vector of length \(M\)) \\
                   & \(y \equiv (D_1,D_2,\ldots,D_{n_p})\) where \(D_p\) are the data points \\
                   & accumulated during exposure \(p\). \\
\(x\)              & solution of the equation \(H x= y\) (vector of length \(N\) ). \\
                   & Also time series (`image space') \(x \equiv x_{1:\Nseries}=(x_1,\ldots,x_{\Nseries})\) \\
                   & where \(x_i\) are supposed sequentially ordered in time \\
\(H\)              & real rectangular matrix of size \(M \times N\)       \\
\(H^{(J)}\)        & the submatrix (columns of \(H\)) related to source named or \\
                   &numbered J, such that \(H \equiv  [H^{(1)},\cdots,H^{(N)}]\) \\
\(A\)              & square real symmetric matrix of order \(N\)  (\(A=H^TH\))  \\

\(\varepsilon\)   &  measurement errors on x (time series vector of length \Nseries)  \\
                  & assumed to have zero mean, to be independent and normally \\
                  & distributed with a known variance \(\Sigma\)  ( \(\epsilon_i \sim N(0,[\sigma_i^2]) \) ) \\

\(T\)             & \(t_i\) is the time where the data point \(x_i\) is measured with  \\
                  & measurement error \(\epsilon_i\). \(T=(t_1,\ldots,t_{\Nseries})\). \\
\(f(t_i)\)        & the model of a series, this is a piecewise constant function. \\
\(m\)             & number of change points that defined the \(m+1\) segments such  \\
                  & that \(f(t)\) is constant between two successive change points \\
\(n_{seg}(J)\)    & number of segments to describe source named or numbered J \\
\(\tau\)          &  change points such that \(\tau_0=\min(T)\) and \(\tau_{m+1}=\max(T)\) \\
                  & or  in point number units, \(\tau_0=1\) and \(\tau_{m+1}=\Nseries+1\) \\
                  & (\(\tau_0 < \tau_1 < \ldots < \tau_{m+1}\)).\(\tau^*\) denotes the last found change point \\
\(s\)             & \(s_k\) is the value of the model \(f(t)\) for the segment defined by \\
                  & \(\tau_{k-1} \leq t <  \tau_{k}\) (vector of length \(m+1\) ) \\
\(F(\tau)\)       & recursive function to be minimized with respect to \(\tau\) \\
\(\beta\)         & penalty to prevent  overfitting \\
\(x_{(\tau_{i-1}+1):\tau_i}\) &  subset of the vector \(x\), \(x_{(\tau_{i-1}+1):\tau_i}=(x_{\tau_{i-1}+1},\ldots, x_{\tau_{i}})\) \\
\( \mathcal{C} (x_{(\tau_{i-1}+1):\tau_i}) \) & Cost function or effective fitness for the   \\
                  & vector \(x\) subset \(x_{(\tau_{i-1}+1):\tau_i}\) \\
\(n\)             & Iteration number which corresponds to time \(t_n\) \\
\end{tabular}
\end{tiny}
\tablefoot{Initial state is denoted with the upper subscript \(^0\): (\(H^0, N^0, \ldots\)).
          The temporary state, with the upper subscript \(^*\): (\(H^*, x^*, N^*, n_{seg}^{*}, \ldots\)).
          The time \(t\) is sometimes replaced by the iteration number \(n\).}
\label{table:indexmath}
\end{table}
\section{Theory and calculation}
\label{sec:segmentation}
%
\subsection{The `image-space' algorithm - partition of a time series}
\label{sec:segmentation:image}
The partition of an interval into segments is closely related to the
topic of change points, which is widely discussed in the literature. There is
a variety of efficient ways to analyze a time series if the parameters
associated with each segment are independent \citep[][and references
therein]{Fearnhead06,Hutter07}.  
\citet{Scargle98} proposed an algorithm for best data partitioning
within an interval based on Bayesian statistics. 
Applications to
astronomical time series (BATSE bursts characterization) can be found
in \citet{Scargle98}.
%
\subsection{Problem formulation}
\label{sec:segmentation:theory}
A list of notations used throughout this paper is given in Table ~\ref{table:indexmath}.
We consider the time series  
\(x \equiv x_{1:\Nseries}=(x_1,\ldots,x_{\Nseries})\), comprising \Nseries\ sequential elements,
following the model
\begin{equation}
    x_i \equiv f(t_i)+\epsilon_i  \mathrm{~~~} i=1,2,\ldots,\Nseries,  \label{eqn:series}
\end{equation}
where \(x_i\) are the measured data and \(\epsilon_i\) their
measurement errors. 
The data are assumed to be ordered in time, although they may be spaced irregularly,
meaning that each \(x_i\) is associated
with a time \(t_i\) and contained in a time interval \(T=(t_1,\ldots,t_{\Nseries})\). 
\(f(t_i)\) is the model to be determined.  
We chose to model the time series
as a combination of piecewise constant segments of time or blocks. 
This set of non-overlapping blocks that add up to form the whole
interval forms a partition of the interval \(T\).
Hence there are, \(m\) change points \(\tau_{1:m}=( \tau_1,
\ldots, \tau_m)\), which define \(m+1\) segments, such that the function
\(f(t)\) is constant between two successive change points,
\begin{equation}
f=\sum_{k=1}^{m+1} s_k\mathcal{I}_k \mathrm{~~with~~} \begin{cases}\mathcal{I}_k=1&\text{if }t \in [\tau_{k-1},\tau_k[~\\\mathcal{I}_k=0&\text{otherwise,}\end{cases}  \label{eqn:compact}
\end{equation}
Here \(\tau_0=\min(T)\) and \(\tau_{m+1}=\max(T)\) or, equivalently, in
point number units, \(\tau_0=1\) and \(\tau_{m+1}=\Nseries+1\) (\(\tau_0 < \tau_1 < \ldots < \tau_{m+1}\)).
Since these data are always corrupted by observational errors, the aim
is to distinguish statistically significant variations through the
unavoidable random noise fluctuations.  Hence, the problem is 
fitting a function through a noisy one-dimensional series, where
the function \(f\) to be recovered is assumed to be a piecewise constant
function of unknown segment numbers and locations (see 
Appendix~\ref{app:phystomath}).
%
\subsection{Algorithm}
\label{sec:segmentation:alg}
\begin{figure}[!ht]
\centering
\includegraphics[width=0.5\textwidth]{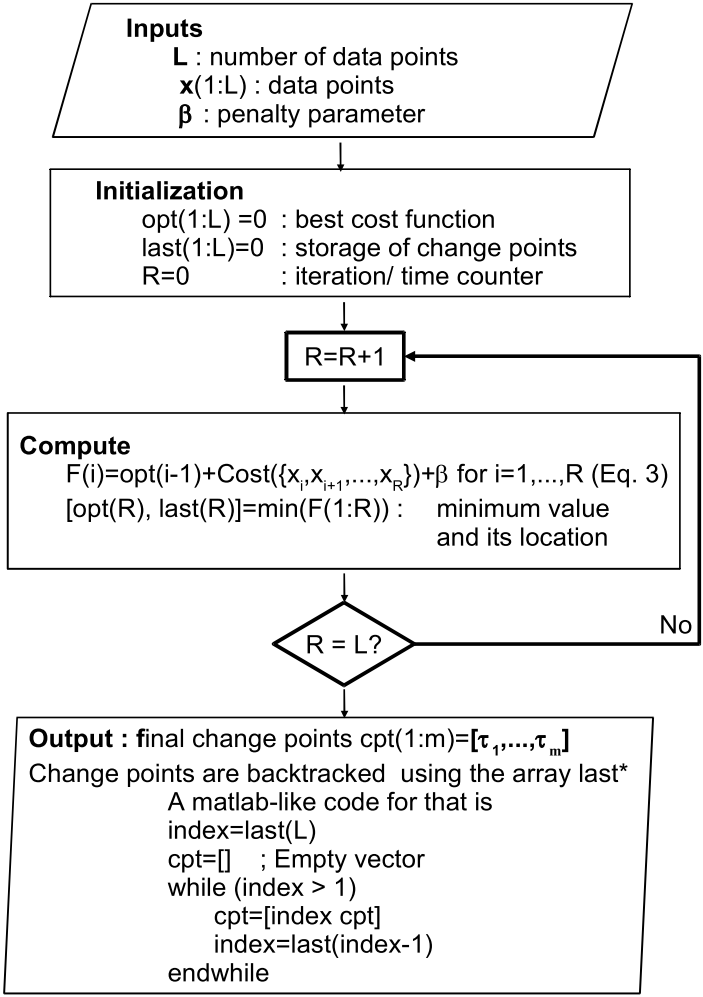}
\caption{Flowchart of the `image-space' algorithm (pseudo-code in Appendix ~\ref{app:imagespace:algo}). The vector  \(last\) gives the starting point of 
individual segments of the best partition in the following way; Let \(\tau_{m}\)=\(last(\Nseries)\), \(\tau_{m-1}\)=\(last\)(\(\tau_{m}-1\)) and so on. The 
last segments contains cells (data points)  \(\tau_{m},\tau_{m}+1,\ldots,n\), the next-to-last contains cells \(\tau_{m-1},\tau_{m-1}+1,\ldots,\tau_{m}-1\), and so on.}
\label{fig:flowchart_imagespace}
\end{figure}

\citet{Yao84}
and \citet{Jackson05} proposed a dynamic programing algorithm to
explore these partitions.  The algorithm described in
\citet{Jackson05} finds the exact global optimum for any block
additive fitness function (additive independent segments) and
determines the number and location of the necessary segments. For \Nseries\ data points, there are \(2^{\Nseries-1}\) possible partitions. 
In short, these authors proposed a search method that aims at minimizing
the following function (see also \cite{Killick11}):
\begin{equation}
     \min_{{\tau}} \left\{\sum_{i=1}^{m+1}{\left[\mathcal{C}(x_{(\tau_{i-1}+1):\tau_i}) + \beta \right]}\right\}, \label{eqn:objf}
\end{equation}
where \(\mathcal{C}\) represents some cost function.
The \(i\)-th segment
contains the data block \(x_{(\tau_{i-1}+1):\tau_i}=(x_{\tau_{i-1}+1},\ldots, x_{\tau_{i}})\) (in point number
units), and the cost function or effective fitness for this data block is \(\mathcal{C}(x_{(\tau_{i-1}+1):\tau_i}\)).
The negative log-likelihood, for example the chi-square, is a commonly used cost function in the
change point literature, but other cost functions can be used instead
\citep [e.g.][and references therein]{Killick11}.
\(\beta\) is a penalty to prevent  overfitting.
\citet[but see also \citet{Killick11}]{Jackson05} proposed a convenient recursive expression to build the partition in \(L\) passes,
\begin{equation}
	F(n)=\min_{\tau^*}\left\{F(\tau^*)+\mathcal{C}(x_{(\tau^*+1):n})+\beta \right\} \mathrm{~n=1,\ldots,\Nseries}. \label{eqn:recursion}
\end{equation}
This enables calculating the global best segmentation using best segmentations on subsets of the data. Once the best segmentation for the
data subset \(x_{1:\tau^*}\) has been identified, it is used to infer the best segmentation for data \(x_{1:\tau^*+1}\).\\
At each
step of the recursion, the best segmentation up to \(\tau^*\) is stored;
for that, the position of the last change point \(\tau^*\) is recorded in a vector \citep{Jackson05}. 
By  backtracking  this vector, the positions of all change points  that constitute the best segmentation at a given 
time or iteration number can  be recovered.
Figure~\ref{fig:flowchart_imagespace} shows a flowchart of the algorithm.
Our implementation is essentially the same as that of \citet{Jackson05}, but see also \citet{Killick11}.
The best partition of the data is found 
in \(O(\Nseries^2)\) evaluations of the cost function.
%
\section{`Data-space' algorithm}
\label{sec:theory:dataspace}
The \SPI{} data contain the variable signals of several sources. The contribution of each of the sources to the data
through a transfer function is to be retrieved. 
As in Sect.~\ref{sec:segmentation:image} and for each of the sources,  the number and positions of
segments are the parameters to be estimated, but this time,  the estimates are interdependent
because of the nature of the instrument coding.
While many algorithms exist to analyze models where the parameters
associated with each segment are independent of those linked to the other segments, there are only a few
approaches, not of general applicability,  where the parameters are dependent from one segment to another \citep[and references therein]{Punskaya02,Fearnhead11}.
However, 
our problem involves a transfer function, which requires a specific treatment.
%
\subsection{Contribution of the sources to the data}
\label{sec:theory:dataspace:system_of_eqns}
The elements related to source \(J\) are denoted with the upper subscript \(^{(J)}\).
The variable signal of the source numbered \(J\) is represented by \(n_{seg}(J)~(= m^{(J)}+1)\) piecewise constant segments (\timebins{})
delimited by \(0 \leq m^{(J)} \leq n_p-1\) change points (Sec.~\ref{sec:segmentation:theory}).
The response \(H^0(1:M,J)\) , which corresponds to a constant source intensity, is a vector of length \(M\) 
(\(=\sum_{i=1}^{n_p} n_d(i)\)). This vector is expanded into a submatrix with \(n_{seg}(J)\) orthogonal columns.
The signal can be written (see also  Eq.~\ref{eqn:compact})
\begin{equation*}
x^0(J) \rightarrow f^{(J)}(t)=\sum_{k=1}^{m^{(J)}+1} s_k^{(J)}\mathcal{I}_k^{(J)}(t)\mathrm{~with~}
\begin{cases}\mathcal{I}_k^{(J)}(t)=1&\text{if } t \in [\tau_{k-1}^{(J)},\tau_k^{(J)}[\\
\mathcal{I}_k^{(J)}(t)=0&\text{otherwise.}\end{cases}
\end{equation*}
The corresponding predicted contribution to the data, at time t (in units of exposure number), is
\begin{equation*}
\begin{split}
\sum_{k=1}^{n_{seg}(J)} H^{(J)}(t,k) s_k^{(J)} \mathrm{~t=1,\ldots,n_p}.
\end{split}
\end{equation*}
The submatrices \(H^{(J)}(t,k)\) of dimension \(n_d(t) \times n_{seg}(J) \) are derived from the response vector \(H^0(1:M,J)\) as follows:
\begin{equation*}
     H^{(J)}(t,k)\!=\!
     \begin{cases}
         H^0(i,J) & \text{if~} \sum_{q=1}^{\tau_{k-1}^{(J)}} n_d(q)+1 \le i \le \sum_{q=1}^{\tau_{k}^{(J)}} n_d(q) \\
        0                             & \text{otherwise.}
     \end{cases}
\end{equation*}
The response associated with the \(N^0\) sources is obtained by grouping subarrays \(H^{(J)}\) of all sources,
\begin{equation}
 H \equiv  [H^{(1)},\cdots,H^{(N^0)} ]. \label{eqn:gather}\\
\end{equation}
Matrix \(H\) has size \(M \times N\) with \(N=\sum_{J=1}^{N^0} n_{seg}(J)\). 
Therefore, the predicted contribution of the \(N^0\) variable sources to the data is
\begin{equation*}
   \sum_{j=1}^{N} H(i,j) x(j) \mathrm{~} i=1,\ldots,M,
\end{equation*}
where \(x=(s^{(1)}_{1:n_{seg}(1)}, \ldots, s^{(N^0)}_{1:n_{seg}(N^0)}) \) is the flux in the \(N\) \timebins{}
(see also Appendix~\ref{app:phystomath}).
%
\subsection{Algorithm}
\label{sec:theory:dataspace:algorithm}
\begin{figure}[!ht]
\centering
\includegraphics[width=0.5\textwidth]{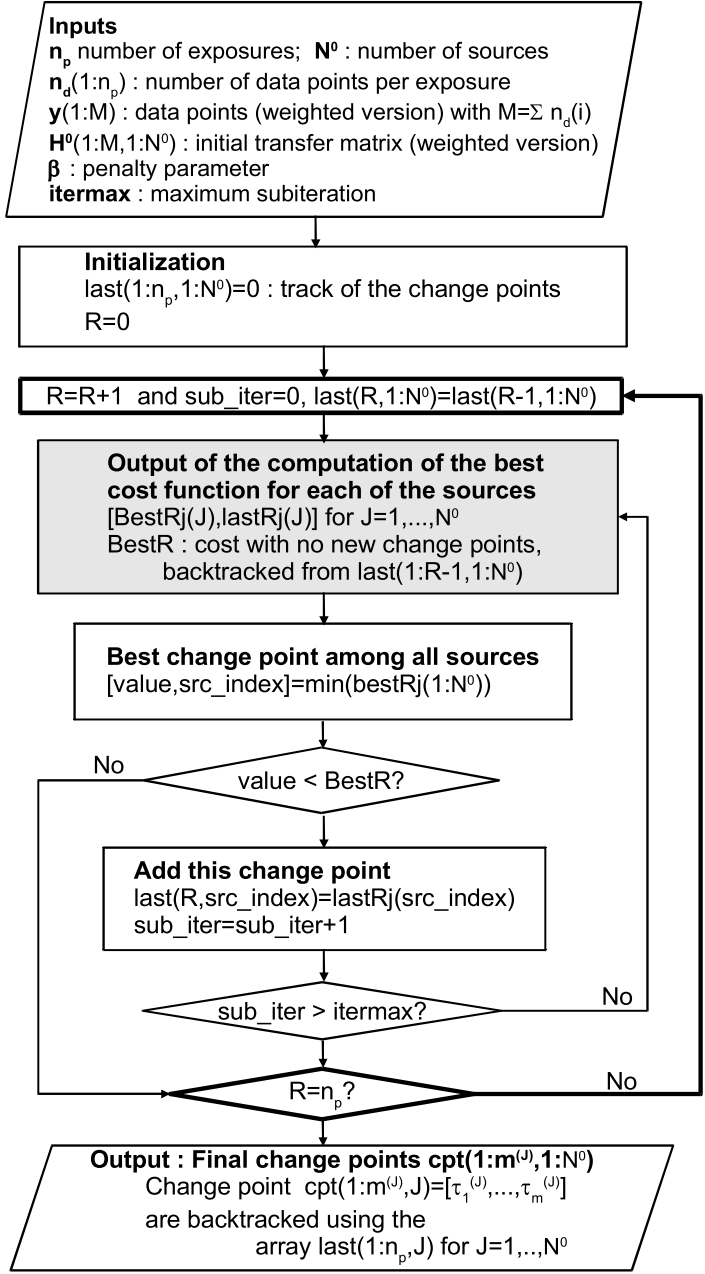}
\caption{Flowchart of the `data-space' algorithm (pseudo-code in Appendix~\ref{app:dataspace:algo}). The cost function calculation, here the shaded box, is detailed in Fig.~\ref{fig:flowchart_datacost}.}
\label{fig:flowchart_dataspace}
\end{figure}

The search path or
the number of possible partitions for \(N^0\) sources and the \(n_p\) exposures is
\((2^{n_p-1})^{N^0}\). An exhaustive search is
not feasible for our application. In addition, computing the cost
function for a particular partition implicitly involves solving a system of equations (computing the cost function),
which renders the search even more inefficient. 
We essentially make a simplification or approximation
that reduces the computation time, allows significant optimizations, and hence makes the problem tractable.\\
Instead of searching for
the exact change points, we can settle for reliable segments and hence
derive an algorithm that aims at significantly reducing the search
path.
Rather than exploring the space for all sources simultaneously, we explore the reduced space associated to a single source at a time. The
algorithm 
closely follows the pattern of Eq.~\ref{eqn:recursion}. It is an iterative construction of
time segments that processes the sources sequentially and independently.  
In practice, the best set of change points using the data recorded  up to time \(t-1\) 
is stored.
The \(n_d(t)\) new data points obtained at time \(t\) are added to the dataset.
The procedure starts for the best change points recorded so far.
For a given source, an additional change point  is added at the various possible positions.
Its best position is recorded  along with the best new cost function.
The operation is repeated for all
sources. The best change point (and cost function value) of all sources is
kept (location and source number) and added to the growing set of
change points. This procedure is repeated several times, for instance \(N_{iter}\) (\(\le N^0\) subiteration)  
times, until no more change points are detected or until the cost function cannot be improved any further.
Then, another iteration at \(t+1\) can start. 
Figure~\ref{fig:flowchart_dataspace} schematically shows the resulting
procedure.\\
This experimental approach is exact when the columns of the
transfer matrix are orthogonal. Although there is no mathematical proof that this holds for a matrix that does not have this property, it is efficient on \SPI\ transfer function and data (Sect.~\ref{sec:results:dataspace}).
%
\subsection{Cost function calculation}
\label{sec:theory:dataspace:cost_function}
The computation of the cost function, used to adjust the data to the model, is detailed in Appendix~\ref{app:segmentation}.
The derived expression involves the minimization of the chi-square function.
Indeed, the cost function must be computed many times,  each time that a new partition is tested. This is by far the most time-consuming part of the algorithm.
Fortunately, these calculations can be optimized. \\
At iteration \(n\), for a subiteration,  we have to solve about \(n \times N^0\) systems
of equations to compute the cost functions corresponding to the different change-point positions.
Instead of solving all these systems, it is possible to
solve only a few of them and/or reduce the number of arithmetic operations  by relying on the structure of the
matrices \footnote{In our implementation, the matrices are sparse and the 
estimated search path number and cost in arithmetic operations are lower than those
given in this section, which correspond to full matrices.}. \\
The matrix \(H\) of size  \(M \times N\) using the change points obtained at iteration \(n-1\) is the starting point.
To search for the best partition for source \(J\), we have to replace only the submatrix \(H^{(J)}\) by its updated version \(H^{*(J)}\).
The corresponding transfer matrix \(H^*\)  (of dimension \(M \times N^*\)) can be written (Eq.~\ref{eqn:gather}) as
\begin{equation*}
 H^*(1:M,1:N^*) \equiv  [H^{(1)},\cdots,H^{(J-1)}, H^{*(J)}, H^{(J+1)},\cdots, H^{(N^0)}].
\end{equation*}
The dimension of  the submatrix \(H^{*(J)}\) corresponding to the new data partition is \( M \times n_{seg}^{*}(J)\).
Here \(n_{seg}^{*}(J)\) is the resulting  number of segments for source \(J\). It varies  from \(1\), which corresponds to a single segment for the  source \(J\), to 
the \(n_{seg}^{max}(J)+1\), where \(n_{seg}^{max}(J)\) is the highest number of segments obtained so far for source J for iteration up to \(n-1\).
In most cases, only the few last columns of the matrix \(H^{*(J)}\) are not identical to those of \(H^{(J)}\), since the very first columns tend to be frozen (Sec.~\ref{sec:theory_dataspace:limited_version}).
If \(n_{seg}^{keep}\) is the number of consecutive identical columns, then
\begin{equation*}
     H^{*(J)} \equiv [H^{(J)}(1:M,1:n_{seg}^{keep} ), H_{add}(1:M,1:N_{add}) ].
\end{equation*}
Here \(H_{add}\) is the modified part of  \(H^{(J)}\).   Let \( k_{min}(J)=\sum_{k=1}^{J-1} n_{seg}(k) \) and  \( k_{max}(J)=\sum_{k=1}^{J} n_{seg}(k)\). Then, the 
columns, from  \(k_{min}(J)+n_{seg}^{keep}+1\) to \(k_{max}(J)\)  are deleted in the matrix \(H\) and \(N_{add}\) columns are added at position
\(k_{min}(J)+n_{seg}^{keep}+1\) to \(k_{min}(J)+n_{seg}^{keep}+N_{add}\).
Therefore, it is not necessary to compute the full matrix \(H^*\)
from scratch each time a new change point is tested (position and source number), but simply to update some  columns of the matrix \(H\).
This construction of the transfer function significantly saves on run time and provides 
clues to further optimization.\\
\begin{figure}[!ht]
\centering
\includegraphics[width=0.5\textwidth]{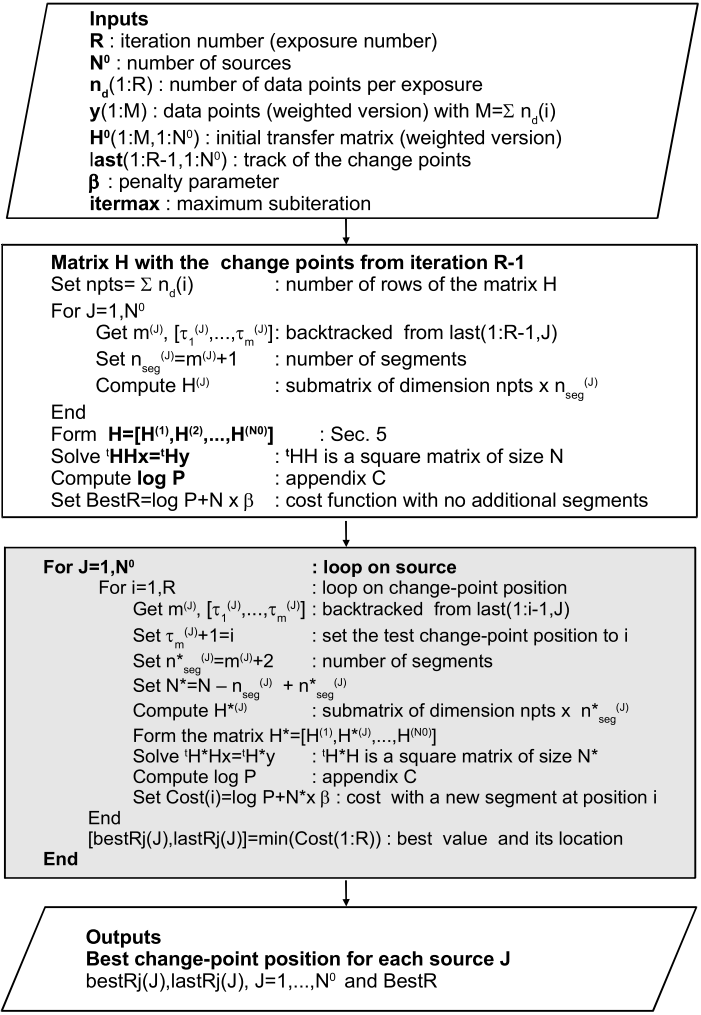}
\caption{Flowchart of the calculation of the cost function for the `data-space' algorithm.}
\label{fig:flowchart_datacost}
\end{figure}
However, in our application, we solve a least-squares system, and hence the matrix
\(A=H^TH\) (\(A^*=H^{*T}H^*\)) is to be implicitly inverted.  We use
an expression based on the Sherman-Morrison-Woodbury formula and the
matrix partition lemma 
(\ref{app:lemna})
to obtain \(A^*\) from \(A\) after subtracting \(N_{sub}\)
and added \(N_{add}\) columns to \(H\) (Appendix~\ref{app:lemna}).
With this optimization, we need only one matrix decomposition (\(A=L^tL\) where \(L\) is a lower triangular matrix)
of a symmetric sparse matrix 
\footnote{We used the MUMPS (MUltifrontal Massively Parallel Solver) package 
(http://mumps.enseeiht.fr/ or http://graal.ens-lyon.fr/MUMPS). The  software provides stable and reliable factors and can process indefinite symmetric matrices. We
used a sequential implementation of MUMPS.}
\(A\) of order \(N\) per iteration and solve the
system (\(A x =H^Ty\)) (store factors and solution).  Then, for each tested
change-point position tested, only one system of equations is to be
solved (using the matrix decomposition stored previously), other
operations are relatively inexpensive. 
To compute the cost function, it can be necessary for our application, depending on the expression of the cost function, to compute a determinant.
However, in
our application, only the ratio of the determinant of the created
matrix (\(A^*\)) of rank \(N^*=N+N_{add}-N_{sub}\)  to matrix \(A\) of rank \(N\)
matters. This ratio is very stable because it involves only matrices of
small rank. In addition, as a cross-product of the update of \(A\), it is
inexpensive to compute. 
%
\subsection{`Limited memory' version}
\label{sec:theory_dataspace:limited_version}
The number of possible partitions, and hence possible change-point positions, increases with the number of exposures
(Sec.~\ref{sec:theory:dataspace:algorithm}). This makes the problem difficult to solve in terms of computing time for data sets containing a large number of exposures.
The positions of the change points are not necessarily fixed until the
last iteration. In the worst case, there can be no common change points (hence segments) between 
two successive iterations.
In practice, it turns out that they become more or less frozen
once a few successive change points have been located. Then, for each
source, instead of exploring the \(n\) possible positions of the
change point, we  consider only the data accumulated between iteration \(n_0 > 1\)  and  \(n\) 
(The 'full' algorithm  starts at \(n_0=1\)). This defines a smaller exploration window size,
which 
further reduces the number of cost function evaluations (in the shaded box of Fig.~\ref{fig:flowchart_datacost}, the loop on change-point positions starts at \(i=n_0 > 1\)). The number of change-point positions to test is nearly  stable as a function of the number of accumulated  exposures.
Section~\ref{sec:results:dataspace:limited_vs_full}) compares the `limited memory' and 'full' version.
%
\subsection{Parallelization}
\label{sec:theory_dataspace:parallelization}
The algorithm sequentially divides each source interval into segments. Hence, it is straightforward to parallelize the code and process several sources simultaneously. To do so, the 
`source loop' (in the shaded box of Fig.~\ref{fig:flowchart_datacost}, the loop on \(J\))
is parallelized using Open-MP. 
Matrix operations are performed in sequential mode, although a parallel implementation is also possible. Experiments on parallelization are
presented in Sect.~\ref{sec:results:dataspace:parallelization}.
%
\section{Results}
\label{sec:results}
%
The sky model for each dataset consists  of sources plus the instrumental background, whose intensity variations are to be determined.
The background is also the  dominant contributor to the observed detector counts or data. Its  spatial structure on the detector plane (relative  count rates of the detectors) is assumed to be known thanks to `empty-field' observations, but its intensity is variable (Appendix~\ref{app:phystomath}).
By default, its intensity variability timescale is  fixed
to \(\sim\) 6 hours, which was shown to be relevant for \SPI\  \citep{Bouchet10}.
Its intensity  variation can also be computed, as for a source, with the `data-space' algorithm.\\
Owing to observational constraints (solar panel orientation toward the Sun, etc.), a given source is  `visible' only every \(\sim\)6 months and is observed only during specific dedicated exposures (Fig.~\ref{fig:lc_gx339_image}). 
For these reasons, it is more convenient to present the `light curves' as a function of the exposure number, so the abscissa does not actually reflect 
the time directly .\\
The fluxes obtained with \SPI{} and \IBIS{} in each defined `time bin', are also compared whenever possible. 
However,  \IBIS{}  light curves are provided by a `quick-look analysis' and are intended to give an approximate measurement of the source intensity.
These time series, used as external data input to the 'image-space' method, 
are not fully converted into units that are independent of the instrument (such as \(photons. cm ^ 2. s ^ {-1}\)) and can even be affected by the specific characteristics of the instrument. 
%
\subsection{Datasets/training data} 
\label{sec:datasets}
%
The various datasets used in this paper are shorter subsets  related to a given sky region of a larger database that consists of \(\sim\)39\,000 exposures covering six years of observations of the entire sky \citep{Bouchet11}.
An \SPI{} dataset, related to a particular source, consists of all exposures whose angular distance, between the telescope pointing axis and the source of interest direction (central source), is less than  \(15^{\circ}\).
This procedure gathers the maximum number of exposures containing the signal from the source of interest, but at the same time the datasets span  a  \(\sim 30^{\circ}\) radius FoV sky region containing numerous sources.\\
Concerning the energy bandwidth of the datasets,
the transfer function is computed for the same mean energy. However, because of the spectral evolution
of the source with time, this approximation  introduces some inaccuracies that increase with the energy bandwidth. 
At first glance, it is preferable to use a narrow energy band, but
at the same time,
the S/N of the sources are lower (since the error bars on the fluxes are large enough,
we detected virtually no variation in intensity).
In other words, for a narrower energy band, the reduced chi-square between the data and the model of the sky is lower, but at the same time, the
source intensity variations are less marked.
Finally, as a compromise for our demonstration purpose, we used the  25-50 keV and 27-36 keV energy bands.
%
\subsection{`Image-space' method} 
\label{sec:results:imagespace}
%
\subsubsection{Application framework}
\label{sec:results:imagespace:framework}
We rely on auxiliary information, such as available light curves, to define the  \timebins\ for \SPI. 
These light curves come from the  \IBIS{} `quick-look analysis'\footnote{Provided by the INTEGRAL Science Data Centre (ISDC).}, more precisely, its first detector layer \textit{ISGRI} \citep{Ubertini03}.
The light curves  of the known sources are not all available, but usually the strongest ones are, if not  
always exactly in the requested energy bands, but this is sufficient  to define some reliable \timebins.\\
The `image-space' method consists of two separate steps. The first does not necessarily involve the \SPI\ instrument, while the second
one involves the \SPI\ only.
%
\subsubsection{`Time bins' : segmentation of an existing time series}
\label{sec:results:imagespace:timeseries}
\begin{figure}[!ht]
\begin{center}
\includegraphics[width=0.5\textwidth]{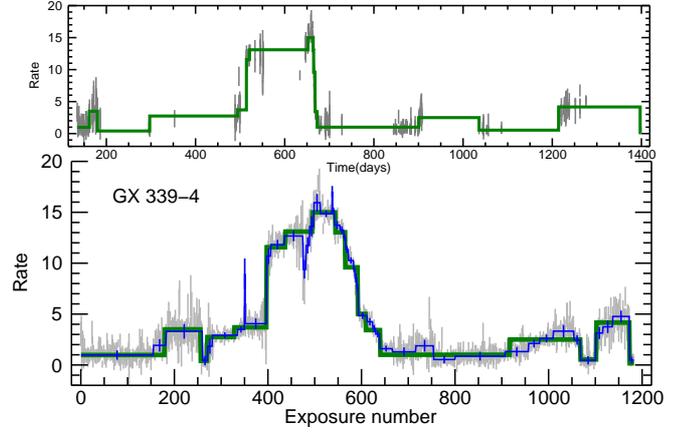}
\end{center}
\caption{GX 339-4 intensity variations. The 26-40 keV \IBIS\  time series (gray), which contains 1183 data points (one measurement per exposure), is segmented into 17 constant segments or \timebins\ (green). The \imagredchi\ between the time series and its segmented version is  1.0006 for 1166 dof. These curves are plotted as a function of time (top)
and the exposure number (bottom). The raw time series (without S/N scaling) is directly segmented into 46 segments (bottom blue curve).}
\label{fig:lc_gx339_image}
\end{figure}

\begin{figure}[!ht]
\begin{center}
\includegraphics[width=0.5\textwidth]{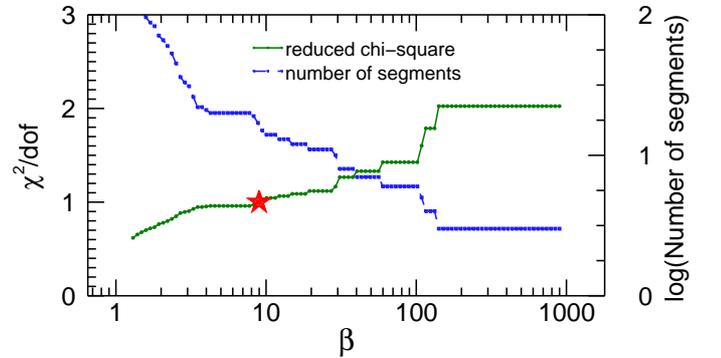}
\end{center}
\caption{GX 339-4 time series segmentation. Variation of the \imagredchi\ (green circles) and number of segments (blue squares) as a function of the penalty parameter value for the time series of fig.~\ref{fig:lc_gx339_image}. The red star indicates the value of the penalty parameter (\(\beta\)=8.99) where  \imagredchi=1.0006 and which corresponds to 17 segments.}
\label{fig:beta_value_gx339}
\end{figure}

The basic process to set up the `time bin' characteristics (position and length) is the time series segmentation.
The raw time series is the light curve linked to a given source, which comes from the `quick-look analysis' of \IBIS\ data. 
Below 100 keV, \IBIS\  is more sensitive than \SPI.
However, it depends on the source position relative to the instrument pointing axis, the number of sources in the FoV, but also on the source's spectral shape. 
The measured S/N ratio of the two instruments for strong sources is around \(\sim 2\) and increased to \(\sim 3\) for crowded sky regions.
To have roughly similar S/N per sources between \IBIS{} and \SPI{}, especially in crowded sky regions, random Gaussian statistical fluctuations are added to raw time series to obtain 
statistical error bars on \IBIS{} light curves increased roughly by a factor \(3\) below 100 keV. This forms the time series used.
This ad hoc procedure is also retrospectively coherent with the number of segments obtained with the 'data-space' method (Sect.~\ref{sec:results:expo_spi_lightcurves}).
Above 150 keV \SPI\ is more sensitive, but  at these energies many sources can be considered to be stable in intensity
because their detection significance is low (fluxes have large error bars), hence we do not expect as many \timebins. \\
In principle, the model of the time series, hence the number of segments, can be adjusted to the data with the desired chi-square value by varying 
the penalty parameter \(\beta\) of Eq.~\ref{eqn:objf}. 
Intuitively, one readily sees that a too high value will underfit, while a too low value will overfit the data.
To be conservative, we adjusted this parameter to obtain a reduced chi-square (\imagredchi) of \(1\); 
the segmentation algorithm is fast enough to test several values of \(\beta\) to achieve this objective.\\
We proceeded as follows: if a single segment (constant flux) gives \imagredchi\  above 1, the time series was segmented.
Many expressions of the cost function can be used; expressions~\ref{eqn:cost_series_marginalized}
and~\ref{eqn:cost_series_flat_prior}  described in Appendix~\ref{app:segmentation:cost} give similar results. Finally, we chose
the most simple one (\ref{eqn:cost_series_marginalized}), which is the chi-square function.\\
Figures~\ref{fig:lc_gx339_image} and~\ref{fig:beta_value_gx339}  show the application to the GX 339-4 source, whose
intensity variation is characterized by a short rise time followed by an exponential decay on a timescale of about one month.
The \IBIS\ time series for the 26-40 keV band contains 1183 exposures. The \imagredchi\ is  1.0006 for 1166
degrees of freedom (dof), the number of \timebins\ is 17 for a penalty parameter value of 8.99 (if the \IBIS\ raw time series is directly segmented
\footnote{The direct segmentation of \IBIS\ time series will generally produce more \timebins\ than necessary. 
This will overfit the \SPI\ model of the source intensity variations.
Obviously, this reduces the chi-square, but does not necessarily produce the best source S/N after \SPI\ data processing.}, the number of segments is 46 instead of 17). \\
%
\subsubsection{Application to \SPI}
\label{sec:results:imagespace:application}
\begin{figure}[!ht]
\begin{center}
\subfigure[GX~339-4]{\includegraphics[width=0.5\textwidth]{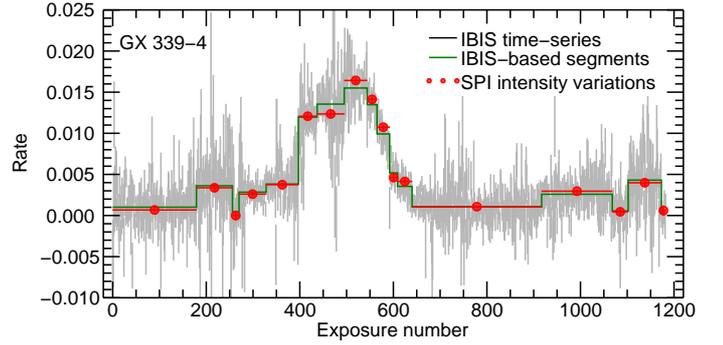} \label{fig:series_GX339}}
\subfigure[GX~1+4]{\includegraphics[trim=0cm 0cm 0cm 0.0cm, clip=true, width=0.5\textwidth]{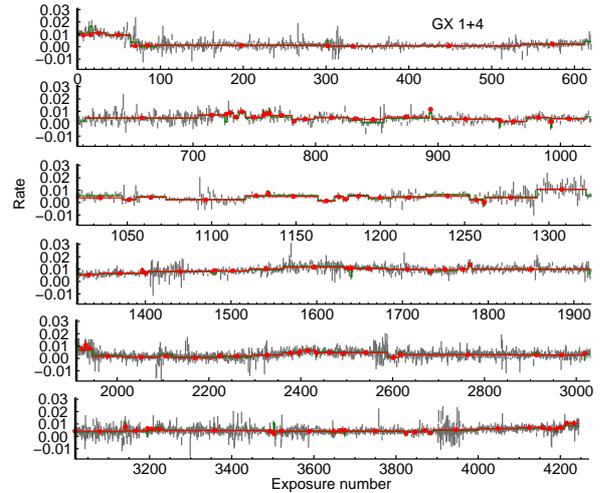}}

\end{center}
\caption{Intensity variations of  GX 339-4 and GX 1+4 in the 27-36 keV band  as a function of the exposure number. 
The \SPI{} segments are plotted in red and the \IBIS{} raw light curves (26-40 keV) are drawn in gray.
The segmented \IBIS{} time series (scaled S/N)  is shown in green. The count rate normalization between \IBIS\ and \SPI\ is arbitrary.
}
\label{fig:imagesegmentation}
\end{figure}

\begin{figure}[!ht]
\begin{center}
\includegraphics[width=0.5\textwidth]{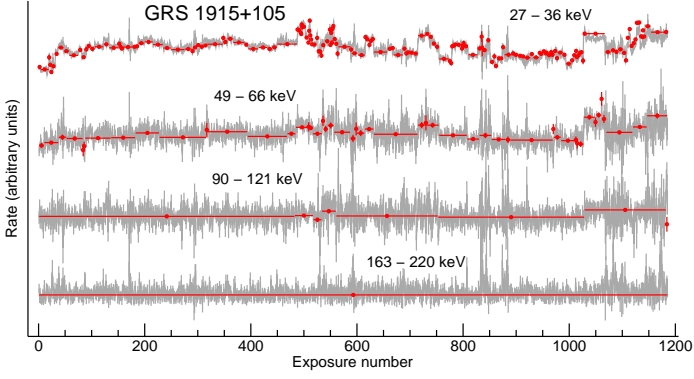}
\end{center}
\caption{`Image-space' method applied to GRS 1915+105. The intensity variations as a function of the exposure number are shown in the 
 27-36, 49-66, 90-121 and  163-220 keV band as indicated in the figure.
The different bands are translated on-to the y-axis (the intensities  are  in arbitrary units).
The \SPI{} segments are plotted in red and the \IBIS{} time series (scaled S/N) are drawn in gray (26-40, 51-63, 105-150, and 150-250 keV).
The number of \timebins{} deduced from the time series segmentation  are 174, 50, 8, and 1.
The global normalization factor in each energy band between the fluxes of 
the two instruments is arbitrary.}
\label{fig:imagesegmentationGRS1915_4bnds}
\end{figure}

\begin{figure}[!ht]
\begin{center}
\includegraphics[trim=0cm 0cm 0cm 0cm, clip=true, width=0.45\textwidth]{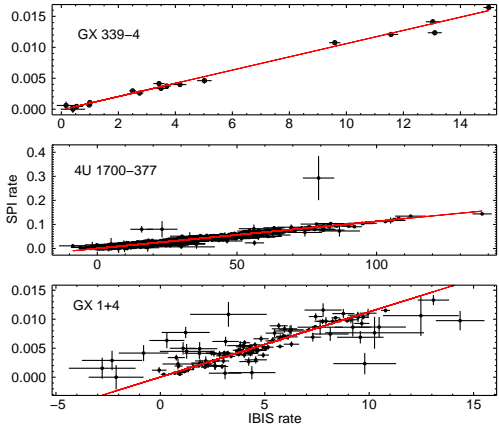}
\end{center}
\caption{\IBIS/\SPI\ flux correlation in 17, 2245, and 122 \timebins\ for GX 339-4, 4U 1700-377, and GX 1+4.
The red line is a straight-line fitting (with errors in both directions) with a chi-square of 25 (15 dof), 3961 (2243 dof), and
321 (120 dof).}
\label{fig:image_flux_correlation}
\end{figure}

We applied the  'image-space' algorithm  to the common \IBIS/\SPI\ 27-36 keV datasets related to the GX 339-4, 4U 1700-377, and GX 1+4 sources, which contain 1183, 4112, and 4246 exposures.
The persistent source 4U 1700-377 is variable on hour timescales,  while GX 1+4 is variable on a timescale from days to  about 1-2 months.
In the sky model are 120, 142, and 140 sources included for the  GX 339-4, 4U 1700-377, and GX 1+4 datasets. \\
For 4U 1700-377, the reduced chi-square (\dataredchi) between the predicted and measured data is 6.15; assuming that all sources have constant intensity, this value is clearly not acceptable.
Each  available \IBIS{} individual light curve was segmented as indicated in sec.~\ref{sec:results:imagespace:timeseries} to define the \timebins.
A total number of 4\,005 \timebins{} was obtained to describe the dataset, and the resulting  4U 1700-377 was partitioned into 2\,245 segments.
The \SPI-related system of equations (Appendix~\ref{app:phystomath}) was then solved using these segments.
The \dataredchi\ is 1.28 for 68455 dof, which is a clear improvement compared with the value of 6.15 obtained when all sources are assumed  to remain constant over time.
It is also possible to directly use the \timebins{} defined with the \IBIS\ light curves without scaling the sensitivities of the two instruments (sec.~\ref{sec:results:imagespace:timeseries}), in which case the \dataredchi\ reaches 1.13.
For strong and highly variable sources such as 4U 1700-377, one can fix the variation timescale to the exposure duration
(\(\sim~1\) hour). All these results are summarized  in Table~\ref{table:imagespace}.\\
For all three datasets, the source 4U 1700-377 was assumed to vary on the exposure timescale. Assuming 
that all other sources are constant in intensity  yields a final \dataredchi\ of 2.46, 1.81, and 2.01 for the 
GX 339-4, GX 1+4, and 4U 1700-377 datasets. \\
The segmentation of the available \IBIS\ time series permits one to obtain better \timebins\ and the \dataredchi\ improves to values below 1.2 (Table~\ref{table:imagespace}).
The resulting GX 339-4 and  GX 1+4 light curves were partitioned into 17 and 122 segments (Table~\ref{table:imagespace}).
Figure~\ref{fig:imagesegmentation} shows the intensity variations of these two sources, obtained from \SPI\ data.\\
The  correlation of the fluxes in the \timebins\ of the two instruments is purely indicative since \IBIS\ time series are obtained from a `quick-look analysis'.
However, the fluxes are well correlated (fig.~\ref{fig:image_flux_correlation}). The linear correlation coefficients are 0.995, 0.941, and 0.816, for GX 339-4, 4U 1700-377, and GX 1+4.\\
To illustrate the decreasing number of \timebins{} with increasing energies, we used the GRS 1915+105 dataset. There are 1185 common \IBIS{} and \SPI{} exposures for this source in our database. For each energy band, the procedure is the same as before, but with 66 sources contained in the FoV.
The number of \timebins\ after segmentation of the \IBIS\ time series in the 27-36, 49-66, 90-121 and 163-220 keV bands, is
174, 50, 8, and 1  and the \dataredchi\ (dof), after solving the \SPI\ related system of equation,  is 1.20 (20\,257), 1.05(20\,545), 1.02(20\,562), and 1.01(20\,616). Figure~\ref{fig:imagesegmentationGRS1915_4bnds} shows the resulting intensity variations.\\
\begin{table}[!ht]
\caption{`Image-space' method final chi-square.}
\label{table:imagespace}
\centering
\begin{tiny}
\begin{tabular*}{0.5\textwidth}{l cccc}
\hline\hline
Dataset          & 4U 1700-377  &    GX 339-4$^*$    & GX 1+4$^*$     & 4U 1700-377$^*$  \\
Exposures        & 4112         & 1183               & 4246           & 4112             \\
Sources          & 142          & 120                & 140            & 142     \\
\hline
\multicolumn{5}{c}{All sources are assumed to have constant intensity}    \\
\dataredchi(dof) & 6.15(71594)  & 2.46(19308)        & 1.81(69361)    &  2.01(67483)  \\
$n_{seg}$        & 1            &   1                & 1              & 4112  \\         
\multicolumn{5}{c}{'Time bins'  from \IBIS\ light curves after S/N scaling (Sec.~\ref{sec:results:imagespace:timeseries})}    \\
\dataredchi(dof)  & 1.28(68455) & 1.186(18880)      & 1.193(68557) &  1.186(66588)   \\
$n_{seg}$         & 2245         &   17              & 122            &  4112      \\    
\multicolumn{5}{c}{'Time bins' from \IBIS\ light curves directly}    \\
\dataredchi(dof)  & 1.132(65768) & 1.106(18317)      & 1.143(66939)   & 1.124(64841)  \\
$n_{seg}$         & 3185         &   46              & 675            & 4112         \\        
\hline
\end{tabular*} 
\end{tiny}
\tablefoot{$^*$The source 4U 1700-377 is also contained in the FoV and is set to be
variable on the exposure duration timescale (\(\sim\) 1 hour).}
\end{table}
%
\subsection{`Data-space' method}
\label{sec:results:dataspace}
\subsubsection{Application framework}
\label{sec:results:dataspace:framework}
As examples, we built datasets related to the sources V0332+53, Cyg X-1, GRS 1915+105, GX 339-4, and IGR J17464-3213 as
described in Sect.~\ref{sec:results:imagespace}. The characteristics  of these datasets are indicated in Table~\ref{table:table1}.
%
\begin{table}[!ht]
\caption{Characteristics of the datasets.}
\centering
\label{table:table1}
\begin{tabular}{l ccc }
\hline\hline
Dataset      & Number of  & Energy      & Number of \\
             & exposures  & range (keV) & sources   \\
\hline
V0332+53    &  391       &  25- 50       &  17 \\
GX 339-4$^a$ & 1183       &  27- 36       & 120 \\
GRS 1915+105 & 2980       &  27- 36       &  61 \\
Cyg X-1      & 2351       &  27- 36       &  32 \\
IGR J17464-3213 & 7147        &  27-36   &   132        \\
\hline
\end{tabular}
\tablefoot{\(^a\)The dataset is restricted to common \IBIS{} and \SPI{} exposures.}
\end{table}
%
\subsubsection{Cost function, penalty parameter, and chi-square }
\label{sec:results:dataspace:penalty_cost}
\begin{figure}[!ht]
\centering
\includegraphics[trim=0cm 0cm 2cm 0.5cm, clip=true,width=0.48\textwidth]{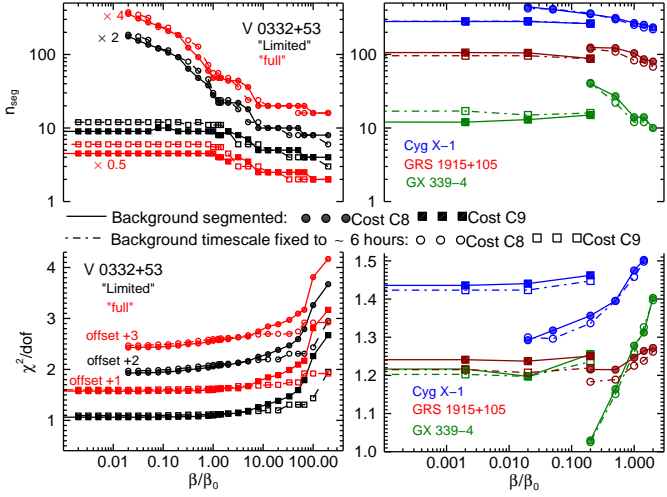}
\caption{(Bottom) Reduced chi-square (\dataredchi) as a function of the penalty parameter in units of \(\beta_0\)=\(\log(M)\), where \(M\) is the number of data points of the dataset. (Top) Number of segments to describe the intensity variation of the central source. Left panels are related to V0332+53  and right panels to GX 339-4, GRS 1915+105, and Cyg X-1. The results obtained with different configurations
of the `limited version' of the 'data-space' method (background timescale fixed to \(\sim\) 6 hours, cost function expression (\ref{eqn:cost_data_marginalized} or ~\ref{eqn:cost_data_flat_prior}) are compared with the default configuration (cost function~\ref{eqn:cost_data_flat_prior} and background timescale to be computed). 
The `full' version  is compared with the `limited' version of the 'data-space' algorithm for the V0332+53 dataset.
}
\label{fig:databetachoice}
\end{figure}

The penalty parameter value  determines the final chi-square value between the data and the sky model, 
but its choice is more problematic than with the 'image-space' algorithm.
Figure~\ref{fig:databetachoice} illustrates the influence of parameter \(\beta\) on the final number of segments and the \dataredchi\ value for 
several datasets of Table~\ref{table:table1}. Because of the very schematic representation of the data, it is not always possible to reach a  \dataredchi\ value of 1
for all these datasets, but 
this value can be approached; if necessary, the width of the energy bands can be reduced (Sec.~\ref {sec:datasets}) to obtain a lower \dataredchi\ value.
Two expressions of the cost function are used using different assumptions (Appendix~\ref{app:segmentation:cost}), but  
there is no strong justification for preferring one expression over the other. \\
The simplest cost function expression~\ref{eqn:cost_data_marginalized} is essentially the chi-square. The 
value of the penalty parameter \(\beta\) prescribed by the Bayesian information criterion is  \(\beta_0 \sim \log (M) \), \( M\) being the number of data points
(Appendix~\ref{app:penalty}).
A penalty parameter value  lower than this prescription  is required to reach a \dataredchi{} value closer to 1. These lower values of \(\beta\) produce a greater number of \timebins, which has the effect of reducing the chi-square at the expense of the reliability of the solution. 
However, the value \(\beta_0\) remains a  reliable guess value of the penalty parameter and gives a number of \timebins\ similar to what is expected from the direct segmentation of the light curves (Sec.~\ref{sec:results:expo_spi_lightcurves}). \\
Expression~\ref{eqn:cost_data_flat_prior} of the cost function has an
additional term  with respect to the expression~\ref{eqn:cost_data_marginalized}, which acts as a penalty parameter and 
further limits the increase of the number of \timebins. 
The minimum cost function value is reached for penalty parameters \(\beta\) below one tenth of \(\beta_0\). A value \(\beta=0\) can be used since 
it gives the minimum chi-square or at least  a value close to the lowest one. The \dataredchi\ value is then comparable to the value obtained with  expression ~\ref{eqn:cost_data_marginalized} and the Bayesian information criterion prescription for the value of \(\beta\).\\
Figure ~\ref{fig:databetachoice} shows the evolution of the \dataredchi\ and total number of segments to describe the intensity variations of the 
central source as a function of the penalty parameter \(\beta\) for several datasets and configurations of the 'data-space' algorithm. 
The \ref{eqn:cost_data_marginalized} expression of the cost function permits one to reach a lower value of the chi-square compared to ~\ref{eqn:cost_data_flat_prior}.
Tables~\ref{table:table2} and ~\ref{table:table3} summarize the results obtained with the 'data-space' method.
We quite arbitrarily chose to use the expression~\ref{eqn:cost_data_flat_prior} as the baseline cost function with penalty parameter \(\beta=0\).
%
\subsubsection{Parallelization}
\label{sec:results:dataspace:parallelization}
Table~\ref{table:dataseg_time_gain} shows experimental results for two  datasets (V0332+53 and GX 339-4 (Table~\ref{table:table1}). A system with 64 (2.62 GHz) cores was used. The run time was measured in the `source loop' (Sect.~\ref{sec:theory_dataspace:parallelization}).
Using 16 threads instead of 1 speeds up the computation by a factor \(\sim\)3 for the V0332+53 dataset and by a factor \(\sim\)4 for the GX 339-4 with the `limited memory' code. The gain is the same for the V0332+53 dataset with the 'full' code, but the average time to perform a loop is about seven times higher than the `limited memory' code. 
\begin{table}[!ht]
\caption{Gain in time achieved by ``source loop'' parallelization of the 'data-space' algorithm.}
\label{table:dataseg_time_gain}
\centering
\begin{tiny}
\begin{tabular}{l cc ccc cc}
\hline\hline
                & Algorithm & \multicolumn{6}{c}{Average time (s) per ``source loop''} \\
Dataset         & version   &  \multicolumn{6}{c}{Number of threads}        \\
                &           & 1     &  2     & 4     &  8    &   16  &  32  \\
\hline
V0332+53       & Full      & 22.8  &  15.7  &  10.1 &  8.1   &  7.7 &  5.2  \\
(391 exposures) & Limited   &  3.1  &   2.0  &  1.8  &  1.1   &  1.0 &  0.7  \\
\hline
GX 339-4         &  Full    &       &        &      & 607   &       &       \\
(1183 exposures) & Limited  & 119   &        &      &  43   &  30   &  23   \\
\hline
\end{tabular} 
\end{tiny}
\tablefoot{The `limited' version is parameterized as indicated in Sec.~\ref{sec:results:dataspace:limited_vs_full}.
The dataset characteristics are detailed in Table~\ref{table:table1}.
A computer with 64 (2.62 GHz) processors was used.
The total number of `source loops' performed for the `full' version was 519 and 2359 for V0332+53 and GX 339-4.
Similarly, 508 and 2296 loops were performed with the `limited memory' version.
The total time spent in the `source loop' for the `full' (`limited memory') 'data-space' algorithm was obtained by multiplying 
the average single-loop time by the number of performed loops.
}
\end{table}
%
\subsubsection{`Limited memory' and `full' version}
\label{sec:results:dataspace:limited_vs_full}
The `limited memory' version (Sect.~\ref{sec:theory_dataspace:limited_version}) is parameterized such that at iteration n, it starts the change-points position exploration at exposure 
\(n_0=\min\)(\(n\) \(-\) 30 exposures, \(n\) - \(2\) change points backward), individually for each source.\\
For the GX 339-4 dataset, the average time spent in a single `source loop'
by the `full' version
increases  with the iteration number (number of exposures processed) while it is comparable per iteration number with 
the `limited memory' version. Table~\ref{table:dataseg_time_loop} reports the gain in time achieved with the `limited memory' version compared with the full `data-space' algorithm.
Finally,  the `limited memory' version  is almost \(\sim\)15 times faster than the `full' version on the GX 339-4 dataset.\\
\begin{table}[!ht]
\caption{
Mean time spent in the `source loop' by the `full' and `limited memory' versions of the 'data-space' algorithm.}
\label{table:dataseg_time_loop}
\centering
\begin{tiny}
\begin{tabular}{l cccc cccc}
\hline\hline
              &\multicolumn{7}{c}{`Source loop' duration (s)} & Total   \\
Iteration     & 20  & 50  & 125 & 250 & 500 & 1000 & 1183       &  time(s) \\    
\hline
Full          & 6   & 18  & 50  & 107  & 288 & 1701  &  2748    & 1432864 \\
`Limited'   & 8   & 18   & 20  & 20  & 30  & 107   &  100    &  98746 \\
\hline
\end{tabular} 
\end{tiny}
\tablefoot{
The characteristics of the dataset related to GX 339-4 are detailed in Table~\ref{table:table1}. The `source loop' was parallelized 
on 8 threads of a 64 processors (2.62 GHz) machine.
The `limited' version was parameterized as follows: \(n_0\)=min(  \(n\) \(-\)30 exposures, \(n\) - two change points backward), where \(n\) is the current exposure, individually obtained for each source as explained in Sect.~\ref{sec:theory_dataspace:limited_version}. }
\end{table}
Table~\ref{table:table3} compares the results of the `limited memory' and `full' version 
for the V0332+53 and GX 339-4 and the GRS 1915+105 and Cyg X-1 datasets. 
The number of segments found by the `limited' version is in general slightly higher than the number found by the `full' version of the code; however, this effect is also due to the particular parametrization used for the `limited' version.
The results remain quite similar in terms of number of segments and reached \dataredchi, showing that the `limited memory' version is a good alternative to  the full `data-space' algorithm, especially for the gain in computation time. 
The gain in time of the `limited' (with the above parametrization) over the `full' version reaches a factor \(\sim\)30 when the background timescale is fixed (not to be segmented).
%
\subsubsection{Background segmentation}
\label{sec:results:dataspace:background_segmentation}
It is no longer necessary to fix the variation timescale of instrumental background to about six hours. The 'data-space' algorithm  models the change in intensity of the background similarly to the other sources of the sky model (Table ~\ref {table:table3}).
If we let the algorithm determine the background intensity variation, it is
modeled with a fewer number of  segments than when the background variation is fixed for a quantitatively comparable chi-square.
This is because of a better localization of the change points. 
The other parameters, such as the number and locations of the source \timebins, stay essentially the same.
%
\subsubsection{Application}
\label{sec:results:dataspace:application}
\begin{figure}[ht]
\begin{center}
\includegraphics[width=0.5\textwidth]{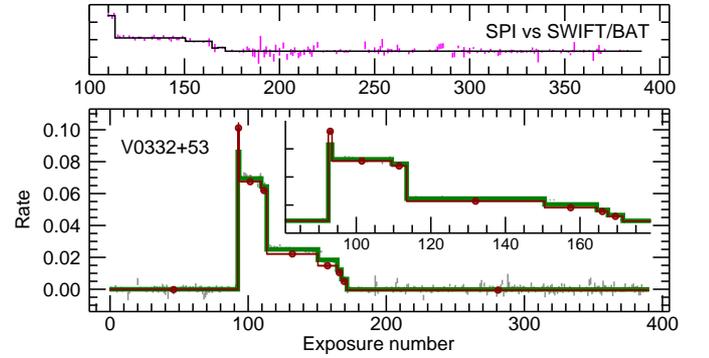}
\end{center}
\caption{
'Data-space' method applied to V0332+53. The source intensity variations (25-50 keV) were modeled by nine segments (red) and were compared with the \IBIS\ time series (26-51 keV, gray). The green curve
corresponds to the \IBIS\ flux averaged on \SPI-obtained segments. The insert is a zoom  between exposure number 81 and 179. (Top) \SPI\ intensity variations model (black) compared with \SWIFTBAT{} time series (24-50 keV, purple line).
The scale between the different instruments is arbitrary and was chosen such that their measured total fluxes are equal.
It should be noted that the \SWIFTBAT{}  and \IBIS{} data are not  necessarily recorded at the same time as the \SPI\ data, nor exactly in the same energy band.}
\label{fig:datasegv0332}
\end{figure}

\begin{figure}[ht]
\centering
\includegraphics[width=0.5\textwidth]{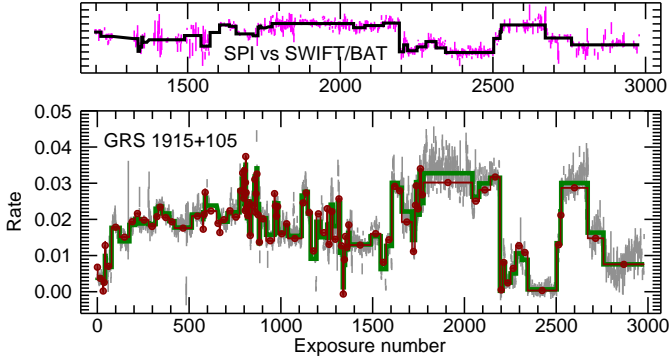}
\caption{Same caption as Fig.~\ref{fig:datasegv0332}, but for GRS 1915+105 in the 27-36 keV band.
The intensity  linear correlation coefficients  
are 0.98 between \IBIS{} (26-40 keV) and and \SPI{}, and 0.97 between \SWIFTBAT{} (24-35 keV) and \SPI.} 
\label{fig:multikill_GRS1915}
\end{figure}

\begin{figure}[ht]
\centering
\includegraphics[trim=0cm 0cm 0cm 0cm, clip=true,width=0.45\textwidth]{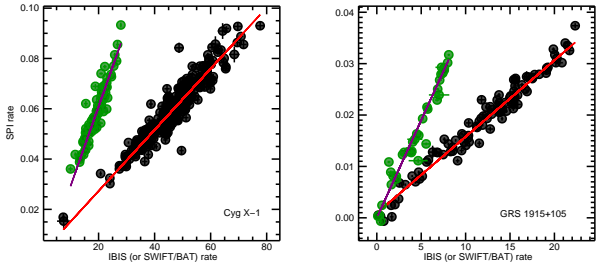} 
\caption{Black denotes the \SPI-\IBIS\ correlation and green the \SPI-\SWIFTBAT\ flux correlation in the 
\timebins\  defined using the
'data-space' method. The best \SPI-\IBIS\ and \SPI-\SWIFTBAT\ linear regression are shown in red and purple.
Left shows Cyg X-1 and right GRS 1915+105.}
\label{fig:multikill_CygX1}
\end{figure}

\begin{figure}[!ht]
\begin{center}
\includegraphics[trim=3cm 0cm 0.5cm 0cm, clip=true, width=0.5\textwidth]{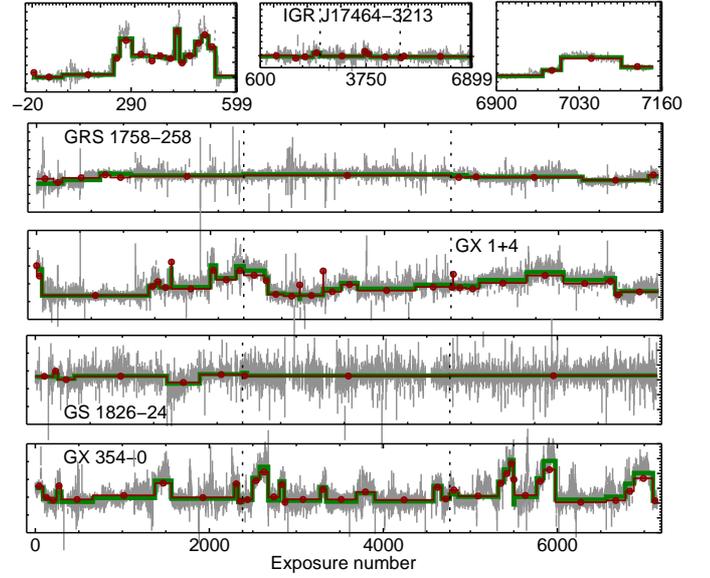}
\end{center}
\caption{Intensity evolutions (red) of IGR~J17464-3213, GRS~1758-258, GX~1+4, GS~1826-24, and GX~354-0 in the 27-36 keV band.
The IGR~J17464-3213 dataset is divided in-to three subsets (dashed vertical lines show the limits) during the computation. 
The \IBIS{} time series (30-40 keV) is shown in gray and the averaged fluxes in \SPI\ \timebins\ are plotted in green, as in  Fig.~\ref{fig:datasegv0332}.
The distance of GRS~1758-258, GX~1+4, GS~1826-24, and GX~354-0 to
IGR~J17464-3213 are 7.3, 8.1, 12.7, and 3.4 \(^{\circ}\).
}
\label{fig:datah1743fov}
\end{figure}

The  `limited memory' version, where the intensity variations of both sources and background are computed, was used as the default algorithm.
For the  dataset relative to V0332+53, this consisted of a \dataredchi\ of 1.06 for 6595 dof and a total of 97 \timebins{}. The resulting V0332+53 intensity evolution is displayed in Fig.~\ref{fig:datasegv0332} as a function of the exposure number (nine segments).\\
Next, we processed the highly variable source Cyg X-1.
The dataset \dataredchi\ is 1.44 for 40\,068 dof and a total of 676 \timebins{}.
The number of segments needed to describe Cyg X-1 is 280.
The relatively high value of the chi-square may be due to the strong intensity and hence to the high S/N of the source. 
However, the systematic errors due to the finite precision of the transfer function start to be important for this strong source, which may be in part responsible
for the high chi-square value.\\
For the dataset related to GRS 1915+105, a moderately strong and variable source, the \dataredchi\ is  1.24 for 51\,573 dof and a total of 440 \timebins{}. GRS 1915+105 intensity variation is displayed in Fig.~\ref{fig:multikill_GRS1915} and consists of 106 segments.\\
The flux linear correlation factors between \SPI\ and \IBIS{} (`quick-look analysis') fluxes measured in the same \timebins{} are 0.991, 0.948, and 0.983 
for V0332+53, Cyg X-1, and GRS 1915+105. The linear correlations with \SWIFTBAT{} are 0.993, 0.934, and 0.973.
The flux correlation for Cyg X-1 and GRS 1915+105 are shown in Fig.~\ref{fig:multikill_CygX1}. 
In addition, the average flux in a given segment is not necessarily based on the same duration of the observation. The number of usable \SPI\ and \IBIS\ exposures contained in the same segment is not always the same and there 
are not always simultaneous observations available
of the same sky region by \SWIFTBAT\ and \INTEGRAL.
Despite these limitations, the fluxes measured by these instruments are quantitatively well correlated.\\
Lower values of the chi-square than those indicated in Table~\ref{table:table3} 
can be obtained by reducing the width of the energy band (Sec.~\ref{sec:datasets}) or by using a lower penalty parameter 
(Sec.~\ref{sec:results:dataspace:penalty_cost}) if
the cost function is given by expression \ref{eqn:cost_data_marginalized}.\\
%
\begin{table}[!ht]
\caption{`Limited-memory' `data-space' algorithm results.}
\centering
\begin{tiny}
\begin{tabular}{l ccc c}
\hline\hline
Dataset        &  Cost        &  Penalty                   &   $\chi^2_r$   & Total  Number    \\
               &  function    &($\beta/\beta_0$)           & (dof)          &  of \timebins    \\
\hline 
      V0332+53 & C.9      &  \textbf{0}                & 1.062(  6595) &   97 \\
                & C.8      &  \textbf{1}                & 1.059(  6592) &  100 \\
                & C.8      &  0.32                      & 0.995(  6492) &  200 \\
                & C.8       &  0.05                      & 0.997( 16533) & 2147 \\
       GX 339-4 & C.9      &  \textbf{0}                & 1.202( 19930) &  940 \\
                & C.9      &  0.02                      & 1.197( 19906) &  964 \\
                & C.8      &  \textbf{1}                & 1.278( 20108) &  762 \\
                & C.8      &  0.2                       & 1.030( 18951) & 1919 \\
   GRS 1915+105 & C.9      &  \textbf{0}                & 1.241( 51553) &  424 \\
                & C.9      &  0.02                      & 1.237( 51565) &  412 \\
                & C.8      &  \textbf{1}                & 1.246( 51585) &  392 \\
                & C.8      &  0.2                       & 1.215( 51023) &  954 \\
        Cyg X-1 & C.9      &  \textbf{0}                & 1.436( 40068) &  676 \\
                & C.8      &  \textbf{1}                & 1.475( 40120) &  624 \\
                & C.8      &  0.02                      & 1.292( 38948) & 1796 \\
\hline
\end{tabular}
\end{tiny}
\tablefoot{The `limited' version is parameterized as indicated in Sect.~\ref{sec:results:dataspace:limited_vs_full}.
The penalty parameter is measured in units  of \(\beta_0\)=\(\log(M)\), where \(M\) is the number of data points of the dataset.
The bold penalty parameters used throughout the analysis are indicated in bold.
\ref{eqn:cost_data_marginalized} indicates that the cost function is the chi-square and
\ref{eqn:cost_data_flat_prior} the cost function obtained with a `flat-prior' hypothesis.}
\label{table:table2}
\end{table}
%
%
The IGR J17464-3213 dataset corresponds to the central, crowded, region of the Galaxy. The sky model consists of 132 sources and the background timescale
was fixed to about six hours.
The dataset is relatively large (7147 exposures) and was artificially split into three subsets to reduce the computation time.
The characteristics of these three subsets are displayed in Table~\ref{table:table2}. The intensity variations of the central source, IGR 17464-3213, was 
modeled with 29 segments. We also extracted the intensity evolutions of GRS 1758-258, GX 1+4, GS 1826-24, and GX 354-0 which are
derived simultaneously (Fig.~\ref{fig:datah1743fov}).
There are a few spikes in the GX 1+4 intensity variations, which are not investigated here in more detail.
The comparison of the source intensity variations obtained with the \IBIS\ instrument (the light curves are assembled from a 'quick-look analysis', the intensities are computed/averaged 
in the temporal segments found by the 'data-space' method)
gives  linear correlation coefficients of 0.995, 0.800, 0.786, 0.845, and 0.901 for
IGR J17464-3213, GRS 1758-258, GX 1+4, GS 1826-24, and GX 354-0. The  comparison with \SWIFTBAT\ gives 0.996, 0.912, 0.909. 1.00, and 0.968 for the same sources.
%
\begin{table*}[!ht]
\caption{`Data-space' algorithm comparison.}
\centering
\begin{tiny}
\begin{tabular}{l c cccc cccc}
\hline\hline
Dataset       &  Version   & \multicolumn{4}{c}{\ref{eqn:cost_data_marginalized}(Chi-square)}   & \multicolumn{4}{c}{\ref{eqn:cost_data_flat_prior} (``Flat'' prior)} \\
              &            & $\chi^2_r$(dof) & \multicolumn{3}{c}{Number of \timebins} & $\chi^2_r$(dof) & \multicolumn{3}{c}{Number of \timebins} \\
              &            &                 & Central source$^a$ & Background$^b$ & Total$^c$   &             & Central source$^a$ & Background$^b$  & Total$^c$  \\
\hline
V0332+53     &  Full        &   1.062(  6603)  & 12   & 53         &  89                    & 1.075(  6612)  &  9    & 51         &  80  \\
              &  Limited     &   1.059(  6592)  & 14   & 50         & 100                    & 1.062(  6595)  &  9    & 50         &  97  \\
              &  Full$^*$    &   1.094(  6545)  & 14   & 96         & 147                    & 1.100(  6550)  & 12    & 96         & 142  \\
              &  Limited$^*$ &   1.089(  6537)  & 15   & 96         & 155                    & 1.094(  6539)  & 12    & 96         & 153  \\
GX 339-4      &              &                 &      &             &                        & 1.218(20097)    &  13   & 71        & 773  \\
              & Limited      & 1.278( 20108)   &  14  &  57         &  762                   & 1.202( 19930)   &  15   & 87        &  940 \\
              &  Limited$^*$ &   1.259( 19932) &  12  & 267         &  938                   & 1.189( 19730)   &  13   & 267       & 1140 \\
GRS 1915+105  & Limited      & 1.246( 51585)   & 100  &  65         &  392                   & 1.241( 51553)   & 102   & 94       &  424  \\
              &  Limited$^*$ & 1.225( 51162)   & 79   & 515         &  791                   & 1.215( 51133)   &  92   & 515      &  840  \\
Cyg X-1       & Limited      & 1.475( 40120)   & 267  &  82         &  624                   & 1.436( 40068)   & 280   & 121       &  676  \\
              &  Limited$^*$ & 1.458( 39798)   & 252  & 448         &  946                   & 1.423( 39729)   & 290   & 448       & 1015  \\
IGR J17464-3213  & Limited$^*$    &            &      &             &                        & 1.263(132486)   & 29   & 1132       &  3307 \\
\hline
\end{tabular}
\end{tiny}
\tablefoot{The algorithm uses the penalty parameter \(\beta=0\) for cost function~\ref{eqn:cost_data_flat_prior} and $\beta=\beta_0$ 
for cost function ~\ref{eqn:cost_data_marginalized} as indicated in bold in Table~\ref{table:table2}.
\(^*\)The background variability timescale is fixed to about six hours. 
\(^a\)The central source from which the processed dataset is named.
\(^b\) Number of segments to describe the background intensity variations and  \(^c\) the total number of segments of the dataset.
}
\label{table:table3}
\end{table*}
%
\subsection{Comparison of the 'image-space' and 'data-space' methods}
\label{sec:results:imagespace_and_dataspace}
\begin{figure}[!ht]
\centering
\includegraphics[width=0.48\textwidth]{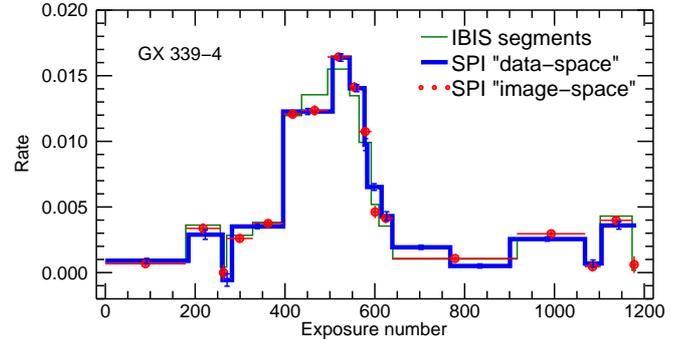}
\caption{Comparison of GX 339-4 (27-36 keV) intensity variations obtained with the 'image-space' and the 'data-space' algorithms.
The common \SPI{}/\IBIS{} database contains 1183 exposures. The `image space'\ method describes GX 339-4 intensity variations
with 17 segments (red) for a \dataredchi\ of 1.19. The 'data-space' method uses 15 segments (blue) and achieves \dataredchi\ of 1.20.
The GX 339-4 segmented version of the \IBIS{} (26-40 keV) time series is shown in green.
 The absolute rate normalization between \SPI{} and \IBIS{} is
arbitrary.}
\label{fig:dataimagcomp}
\end{figure}

The comparison was made on the \SPI{} and \IBIS{} common 1183 exposures related to  the GX 339-4 dataset. 
The `image-space' algorithm used the \IBIS{} (26-40 keV) light curve as external input to define \SPI{} \timebins. 
It  achieved a \dataredchi\ of 1.19 (18880 dof) and a total of 1990 \timebins{} and 
used 17 segments to describe the intensity evolution of GX 339-4 (Sect.~\ref{sec:results:imagespace}). 
The `data-space' algorithm displays
a similar \dataredchi\ of 1.20 and used 15 segments. Both light curves are compared in Fig.~\ref{fig:dataimagcomp}. 
If the background intensity is forced to vary on a timescale of about six hours, the number of segments is 13, if 
the source 4U 1700-377 is forced to vary exposure-by-exposure, the number of segments is 13 as well.
The \SPI\ curves are more difficult to compare quantitatively because the total number of segments and the \dataredchi\ are not the same with the two methods.
(Sec.~\ref{sec:results:expo_spi_lightcurves}). Nevertheless, we compared them indirectly with the \IBIS{} binned ones.
This demonstrates the effectiveness of the `data-space' algorithm.
%
\subsection{Studying a particular source: Obtaining its light curve on the exposure timescale}
\label{sec:results:expo_spi_lightcurves}
\begin{figure}[!ht]
\begin{center}
\includegraphics[width=0.50\textwidth]{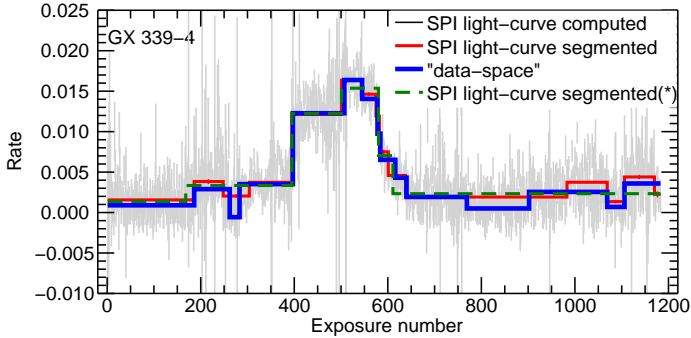}
\end{center}
\caption{`Data-space' algorithm has modeled the temporal evolution of 
GX 339-4 in the 27-36 keV band  with  15 constant segments (blue).
The light curve (in gray) of GX 339-4 was obtained by directly processing  \SPI\ data (after fixing the temporal evolution of 
the other sources).  It contains  1183 data points (one measurement per exposure)
The red  curve is the segmentation of the gray curve into 15 segments with the `image space' algorithm.
(*) If the background is forced to vary exposure-by-exposure, 
the GX 339-4 light curve is segmented only into six segments (dashed-green).}
\label{fig:lc_gx339_spi_image}
\end{figure}

The data related to a given source also contain the contribution of the many other sources in the FoV.
To study this particular source, one  must take into  account  the influence of the other sources and distinguish the different contributions to the data. This means knowing the model of the  intensity variations of all other sources. At the same time,  the number of unknowns to be determined in the related system of equations must be minimum to achieve a high S/N for the source of interest. 
The intensity variation of all sources of the FoV is not generally known a priori and constitutes a difficulty.
The 'image-space' or 'data-space' methods are designed to solve this difficulty.\\
The first step consists of processing data using methods that allow one to construct synthetic intensity variations of the sources; 
then one determines these variations by constructing \timebins\ while minimizing their number. 
The intensity variations of all sources are fixed, excepted those of the source of interest, which can be therefore studied 
in more detail.\\
As an example, the  procedure was applied to the  GX 339-4 light curve study exposure-by-exposure. 
Figure~\ref{fig:lc_gx339_spi_image} shows the light curve obtained with \SPI, compared with its segmented version.
The light curve is segmented into 15 segments, this number is 
comparable to the number found using \IBIS\ time series of 17 segments, which also confirms the idea that it is useful to adjust 
the \SPI\ and \IBIS\ sensitivity.\\
Next, both GX 339-4 source and  background were forced to vary exposure-by-exposure.
The S/N of this GX 339-4 light curve is degraded compared with the case where the  background varies on a timescale of about six hours or is segmented.
The resulting light curve is segmented in-to only six segments.
This illustrates the importance of having a minimum number of parameters to model the  sky and the  background. 
%
\section{Discussion and summary}
\label{sec:discussion_and_summary}
\subsection{Discussion}
\label{sec:discussion_and_summary:discussion}
We proposed two methods for modeling the intensity variations of sky sources and instrumental background. 
The first one, called `image space', is relatively easy to implement and benefits from the \IBIS\ instrument on-board \INTEGRAL, which simultaneously observes the same region of the sky  as \SPI.
Hence in this case, the main weakness of this method is that it requires information from other instruments and then depends both on the characteristics of these instruments (FoV, sensitivity, etc.) and on the level of processing performed on these available external data.\\
The second method, called 'data space'  (Sect.~\ref{sec:theory:dataspace}), is based solely on \SPI{} data and does not depend on external data. With this algorithm, the background intensity variations can be  computed
(\ref{sec:results:dataspace:background_segmentation}).
The dependence across segments through the transfer function
greatly increases the complexity of the algorithm, which is very computer-intensive and hence time-consuming.
We have made some simplifications to be able to handle the problem  and optimizations that in most cases accelerate the computations, in particular the way the matrices are updated during the change-point position search. More or alternative optimizations are possible,
although we did not explore them in this paper; for example a (sparse) Cholesky
update/down-date may be used at each iteration to avoid the decomposition of matrix \(H\) at each iteration. Similarly, an alternative to the
procedure used here could be solving the orthogonal-least-squares problem through \(QR\) rank update.
Because of the approximations we made, the best segmentation (in the mathematical sense) is probably not achieved, but at least we obtained a
reliable one, which is sufficient to improve the sky model.\\
The final value of the chi-square is not really directly controlled for the `image-space' method. For the 'data space', the value achieved depends mainly on the value of the penalty parameter. We empirically derived some initial guess values. 
The determination of the penalty parameter, even if a good guess of its initial value is possible, and the cost function formulation are difficulties of 
the 'data-space' method, which need more investigation.\\
In addition, the 'data-space' method is more suitable for
exploring the interdependence of the source contributions to the data. It takes into account the co-variance of the parameters during the data reduction process. In contrast, the `image-space' method uses only the variance since the light curves are already the product of another analysis and the covariance information related to the sources in the FoV is lost.\\
We chose to use a very simple representation  of the source intensity variations by using piecewise constant functions.
With this modeling of the time series, the individual segments are disjoint. For the 'data-space' method and for a given (single) source, the columns of the corresponding submatrix are orthogonal.
With a more complex light curve model, this property no longer holds.
However, it is possible to derive a formulation where the intensity variations are continuous, e.g., using piecewise polynomials (the quadratic functions have been studied by \citet{Fearnhead11}.
%
\subsection{Summary}
\label{sec:discussion_and_summary:summary}
With only 19 pixels, the \SPI\ detector does not provide enough data to correctly construct and sample the 
sky image viewed through the aperture.
The dithering technique solves this critical imaging problem by accumulating ``non-redundant'' data on a given sky region,
but at the same time it raises important questions of data reduction and image/data combination 
because of the variability of the sources. \\
We proposed two algorithms that model the intensity variation of sources in the form of combinations of piecewise segments of time  during which a given source exhibits a constant intensity. 
For our purposes, these algorithms can help to solve a specific difficulty 
of the \SPI{} data processing, which is to take into account the variability
of sources during observations
and  consequently optimize the signal-to-noise ratio of the sources.
A first algorithm uses existing time series  to build segments of time during which a given source exhibits a constant intensity. This auxiliary information is incorporated into the \SPI\ system of equations to be solved. A second algorithm determines these segments 
directly using \SPI{} data and constructs some ``synthetic'' light curves of the sources contained in the FoV. It separates the contribution to the data of each of the sources through the transfer function. 
Both algorithms allow one to introduce  more objective parameters, here the \timebins, in the problem to be solved.
They deliver an improved sky model that fits the data better.\\
The algorithms also have various applications, for example for studying a particular source in a very crowded region.
One typical example is the extraction of the diffuse emission components in addition to all source components. Indeed, the contribution of variable sources should be `eliminated' across the whole sky, so that the diffuse low surface brightness emission  is as free as possible from residual emission coming from the point sources.\\
In addition, the `image-space' method allows one to merge databases of the various instruments on-board the \INTEGRAL{} observatory and uses the complementary information of  different telescopes. 
%
\section*{Acknowledgments}
The \INTEGRALSPI{} project has been completed under the responsibility and leadership of CNES. 
We are thankful to ASI, CEA, CNES, DLR, ESA, INTA, NASA, and OSTC for their support.
We thank the referees for their very useful comments, which helped us to strengthen and improve this paper. 
%

%
\newpage
\Online
\begin{appendix}
%
%
\section{Stating the problem: from the physical phenomenon to the mathematical equation}
\label{app:phystomath}
The signal (counts and energies) recorded by the \SPI{} camera on the 19
Ge detectors is composed of contributions from each source (point-like
or extended) in the FoV convolved with the instrument
response plus the background.  For \(n_s\) sources located in the FoV, 
the data \(D^{raw}_{dp}\) obtained during an exposure \(p\) in
detector \(d\) for a given energy band can be expressed by the relation
\begin{equation}
D^{raw}_{dp}=\sum_{j=1}^{n_s} R_{dp,j} I^{s}_{p,j} + B^{bg}_{dp} +\epsilon_{dp}, \label{eqn:expression}
\end{equation}
where \(R_{dp,j}\) is the response of the instrument for source \(j\)
(function of the incident angle of the source relative to pointing
axis), \(I^{s}_{p,j}\) is the flux of source \(j\) during exposure \(p\) and
\(B^{bg}_{dp}\) the background both recorded during the exposure \(p\) for
detector \(d\).
\(\epsilon_{dp}\)  are the  measurement errors on the data \(D^{raw}_{dp}\), they are
assumed to have zero mean, to be independent, and to be normally distributed with a known variance \(\sigma_{dp}\)
(\(\epsilon_{dp} \sim N(0,[\sigma_{dp}^2]) \) and \(\epsilon_{dp} = \sqrt{D^{raw}_{dp}}\) ). \\
For a given exposure \(p\), \(D^{raw}_{dp}\),\(\epsilon_{dp}\), \(R_{dp,j}\), and \(B^{bg}_{dp}\)
are vectors of \(n_d(p)\) elements (say, \(n_d=19\)) elements).
However, \(n_d(p)\) corresponds to the number of functioning detectors, the possible values
are 16, 17, 18 or 19 for single events and up to 141 when all the multiple events are used in addition to the single events \citep{Roques03}. 
For a given set of \(n_p\)
exposures, we have to solve a system of \(M\) ( \(=\sum_{p=1}^{n_p}\)) equations
(Eq.~\ref{eqn:expression}).\\
To reduce the number of free parameters related to the background, we take
advantage of the stability of relative count rates between detectors
and rewrite the background term as
\begin{equation}
B_{dp}^{bg} = I^{bg}_p \times U^d \times t_{dp}.  \label{eqn:backmodel}
\end{equation}
In this expression, either \(U^d\) or \(I^{bg}_p\) is supposed to be known,
depending on the hypotheses made on the background structure. To
illustrate the idea, suppose that \(U^d\) is known. 
\(U\) is an `empty field' \citep{Bouchet10}, a vector of \(n_d\) elements.
Then, the  number of parameters necessary to model the background reduces to \(n_p\).
The number of unknowns (free parameters) of the set of \(M\)
equations is then \((n_s+1) \times n_p\) (for the \(n_s\) sources and
the background intensities, namely  \(I^{s}\) and \(I^{bg}\)).  Fortunately, a further reduction of the
number of parameters can be obtained since many sources vary on time
scales longer than the exposure timescale. In addition, many point
sources are weak enough to be considered as having constant flux
within the statistical errors, especially for the higher energies.  Then the \(n_p
\times n_s\) parameters related to sources will reduce to \(N_s^{eff}\)
parameters and, similarly, \(N_b\) for the background.
As these parameters also have a temporal connotation, they are also termed, hereafter, \timebins.
Finally, we have \(N=N_s^{eff} + N_b\)  free parameters to be determined. 
If the source and background relative intensities are supposed to be constant throughout all observations (time spanned by the exposures), 
omitting the detector index, the equation can be simplified as
\begin{equation}
\begin{split}
& D^{raw}_p=\sum_{j=1}^{n_s} R_{p,j)} I^{s}_j +  P_p I^{bg} +\epsilon_p\\
\mathrm{with~}& P_p=t_{d,p}\times U^d.
\end{split}
\end{equation}
Here  \(P_{p}\) is a   vector of \(n_d(p)\) elements.
The aim is to compute the intensities \(I^{s}(j)\) of the \(n_s\) sources  and the background relative intensity \(I^{bg}\).
Therefore, the problem can be written in matrix form as
\[
\left( {\begin{array}{cccc}
 P(1)  &  R(1,1) & ...  & R(1,n_s)  \\
 P(2)  & \ddots  & \ddots  & R(2,n_s)  \\
\vdots & \ddots  & \ddots  & \vdots    \\
P(n_p) & ...  & ...   & R(n_p,n_s) 
 \end{array} } \right)
\left( {\begin{array}{c}
 I^{bg}    \\
 I^s(1)  \\
 I^s(2) \\
 \vdots \\
 I^s(n_s) 
 \end{array} } \right)
=
\left( {\begin{array}{c}
 D(1)   \\
 D(2)  \\
\vdots \\
D(n_p -1) \\
D(n_p)
 \end{array} } \right).
\]
At last, Eq.~\ref{eqn:expression} reduces
to the general linear system \(H^0 x^0 +\varepsilon=y\), where \(H_0\) is
the transfer matrix, \(y\) the data, \(\varepsilon\) the measurement errors and \(x_0\) the unknowns
\(H^0\) (elements \(H_{ij}\)) is an 
\(M \times N_0\) matrix and \(N_0 = n_s+1\) .
\(x^0=(I^{bg},I^{s}(1), \cdots, I^{s}(n_s))\) is a vector of length \(N_0\).
\(y=(D_1,D_2, \cdots, D_{n_p})\)  and \(\varepsilon=(\epsilon_1,\epsilon_2, \cdots, ,\epsilon_{n_p}\)are vectors of length
\(N_0\) and \(M\).
Now, if all \(N_0\) sources (here sources and background) are assumed to vary and the signal of source \(J\) is represented 
by \(n_{seg}(J)\) piecewise constant segments of time or \timebins, then each columns of \(H_0\) will be expanded in \(n_{seg}(J)\) columns.
Eq.~\ref{eqn:expression} reduces to \(H x+\varepsilon=y\), where the matrix \(H\) is of dimension \(M \times N\), with \(N=\sum_{j=1}^{N_0} n_{seg}(j) \)and \(x\) a vector of length \(N\) representing the intensity in the \timebins.  
Schematically and to illustrate the purpose of Sec.~\ref{sec:theory:dataspace:system_of_eqns}, \(H\) can be derived from \(H_0\) :
\begin{small}
\begin{equation*}
\begin{split}
& H^0=
\left( {\arraycolsep=0.5mm\begin{array}{ccccc}
 H_{11}  & H_{12}  & H_{13} & ...    & H_{1N}  \\
 H_{21}  & H_{22}  & H_{23} & \ddots & H_{2N}  \\
 H_{31}  & H_{32}  & H_{33} & \ddots & H_{3N}  \\
 \vdots  & \ddots  & \ddots & \ddots & \vdots \\
 H_{M1}  & H_{M1}  & H_{M3} &...    & H_{MN}  
 \end{array} } \right)\\
\longmapsto &
H= \left( {\arraycolsep=0.5mm\begin{array}{cccccccccccc}
H_{11} & 0      & 0      &   0    &  H_{12}  & 0      & 0      & H_{13} & 0      & H_{1N} & ...    & 0      \\
0      & H_{21} & 0      &   0    &  H_{22}  & 0      & 0      & 0      & H_{23} & H_{2N} & ...    & 0      \\
0      & 0      & H_{31} &   0    &  0       & H_{32} & 0      & 0      & H_{33} & 0      & ...    & H_{3N} \\
\vdots & \vdots & \vdots & \vdots & \vdots   & \vdots & \vdots & \vdots & \vdots & \vdots & \ddots & \vdots \\
0      & 0      & 0      & H_{M1} &  0       & 0      & H_{M2} & 0      & H_{M3} & 0      & ...    & H_{MN}   
 \end{array} } \right).
\end{split}
\end{equation*}
\end{small}
For the training data given in Sect.~\ref{sec:instrument}, which contain \(\simeq 39\,000\) exposures and
for the 25-50 keV energy band, we used~ \( N=N_s^{eff}+ N_b \simeq 22\,500\) \timebins{} \citep{Bouchet11}. 
To fit the sky model consisting of sources plus instrumental background to the data, we minimized the corresponding 
chi-square value:
\begin{equation}
\chi^2=\sum_{i=1}^{M} \left[ \frac{y_i-\sum_{j=1}^{N} H_{ij}x_j} {\sigma_i} \right]^2,
\end{equation}
where  \(\sigma_i\) is the
measurement error (standard deviation) corresponding to the data point \(y_i\), formally \(\sigma_i = \sqrt {y_i}\).
By replacing \(H\) and \(y\) by their weighted version,
\begin{equation}
    y_i \equiv \frac{y_i}{\sigma_i} \mathrm{~and~} H_{ij} \equiv \frac{H_{ij}}{\sigma_i}.
\end{equation}
The least-squares solution 
\(x\) is obtained by solving the following normal equation:
\begin{equation}
(H^T H)x=H^T y \mathrm{~or~as~} A.x=b.
\end{equation}
Here \(A\) is a symmetric matrix of order \(N\), \(x\) and \(b\) are  vectors of length \(N\).
%
\section{Segmentation/partition of a time series}
\label{app:segmentation}
\subsection{Bayesian model comparison}
\label{app:segmentation:bayes}
The literature about model comparisons is vast, especially in the
field of econometric \citep[e.g][]{Zellner71,Poirier95,Koop03}.
We have some data \(y\), a model \(Model_i\) containing a set of parameters
\(\theta_i\).
In our application, we have to choose between models of light curves
(\(Model_1\) or \(Model_2\)), based on observations over a time interval \(T\):
\begin{equation*}
     \begin{cases}
        Model_1: & \text{Constant intensity over T} \\
        Model_2: & \text{Possibly different constant intensities}\\
             & \text{in two subintervals}, T_1+T_2=T.
     \end{cases}
\end{equation*}
Using Bayesian inference (see for example \citet{Scargle98}), 
the posterior odds ratio makes it possible to compare models \(i\) and \(j\):
it is simply the ratio of their posterior model probabilities:
\begin{equation}
PO_{ij}=\frac{p(Model_i | y )}{p(Model_j | y)}= \frac {p(y | Model_i) p(Model_i)} {p(y | Model_j) p(Model_j)}=BF_{ij} \frac{p(Model_i)}{p(Model_j)},
\end{equation}
where \(BF_{ij}\) is the Bayes factor comparing \(Model_i\) to \(Model_j\).
The quantity \(p(y | Model_i)\), the marginal likelihood 
\begin{equation}
p(y | Model_i) \equiv L(Model_i,y)  = \int {P(y | \theta_i, Model_i) P (\theta_i | Model_i) d \theta_i}.
\end{equation}
To compare \(Model_2\) with \(Model_1\), non-informative choices are commonly made for the prior
model probabilities
(i.e, \(P(Model_1)=P(Model_2)=\frac{1}{2}\)).
Then, to prevent overfitting, we introduce a penalty parameter
\(\gamma\) that chooses model \(Model_2\) instead of \(Model_1\) if
\begin{equation}
BF_{21} = \frac{p(y | Model_2)} {p(y | Model_1)} > \gamma.
\end{equation}
To fit the framework of expression Eq.~\ref{eqn:objf}, this
equation can be equivalently rewritten as
\begin{equation}
-\log [p(y | Model_2)] + \beta < -\log [p(y | Model_1)]   \mathrm{~~ with~~} \beta=\log( \gamma).
\end{equation}
%
\section{Cost or likelihood function computation}
\label{app:segmentation:cost}
\subsection{Normal distributions}
The model for the time series observations (Eq.~\ref{eqn:series}) is
\begin{equation*}
x_i \equiv f(t_i)+\epsilon_i \mathrm{~~~} i=1,2,\ldots,\Nseries,
\end{equation*}
where \(x_i\) is the value measured at time \(t_i\), \(f\) is the unknown
signal and \(\epsilon_i\) the measurement errors. These errors are
assumed to have zero mean, to be independent, and to be normally
distributed with a known variance \(\sigma_i^2\), thus the probability
for bin \(i\) is
\begin{equation*}
P(\epsilon_i | \sigma_i)= \frac{1}{\sigma_i \sqrt{2 \pi}} e^{-\frac{1}{2} \left( \frac{\epsilon_i}{\sigma_i} \right)^2}.
\end{equation*}
The parameters associated with each segment are independent of each
other, the likelihood can be calculated for the data within each
segment or block. For the block \(K\) where the true signal is \(\lambda\), 
the likelihood is,
\begin{equation}
\Likeh^{(K)}= \prod_{i \in K} \frac{1}{\sigma_i \sqrt{2 \pi}} e^{-\frac{1}{2} \left( \frac{x_i-\lambda}{\sigma_i} \right)^2}.
\end{equation}
The product is over all \(i\) such that \(t_i\) falls within block \(K\), for instance \(\Nseries_K\) points fall into block \(K\).
We can simply maximize the likelihood for instance, \(P_K=\max(\Likeh^{(K)})\); its maximum is  reached for \(\lambda=\lambda_{max}\). With the following notations
\begin{equation*}
\begin{split}
& a_K= \frac{1}{2} \sum_{i \in K} \frac{1}{\sigma_i^2}~~~b_K=-\sum_{i \in K} \frac{x_i}{\sigma_i^2}~~~c_K=\frac{1}{2} \sum_{i \in K} \frac{x_i^2}{\sigma_i^2} \\
& \mathrm{~and~} C_K=\frac{(2\pi)^{-\Nseries_K/2}} {\prod_{i \in K}^{} \sigma_i}.
\end{split}
\end{equation*}
The expression can be written in terms of ordinary least-squares quantities:
\begin{equation*}
\begin{split}
& \lambda_{max}= \frac{ -b_k}{2 a_k} ~\mathrm{~least-squares~solution}\\
& SSE_K=  \sum_{i \in K} \left( \frac{x_i-\lambda_{max}} {\sigma_i} \right)^2 =-2 c_k +\frac{b_k^2}{2 a_k}~\mathrm{~sum~of~square~errors}.
\end{split}
\end{equation*}
The block likelihood  can be rewritten as
\begin{equation}
\log P_K=-\frac{SSE_K}{2} + \log C_k. \label{eqn:max_serie_lkh}  
\end{equation}
Finally, the probability for the entire dataset is 
\begin{equation}
\log P= \log \left[ \prod_K P_K \right]= \sum_K {\log P_K}.  \label{eqn:cost_series_marginalized}
\end{equation}
The choice of the prior  of \(\lambda\) is another question, although many prior probabilities may be possible, we marginalize it by choosing 
the flat unnormalized
prior (\(p(x | Model)\)= constant). If the variable \( \lambda\) is marginalized,
\begin{equation}
P_K \equiv p(y | Model_K)= C_k~ \int_{-\infty}^{+\infty} {  e^{ - \frac {1} {2}  \sum_{i \in K} {(\frac{x_i-\lambda} {\sigma_i})^2  }}  d\lambda }. 
\label{eqn:marg_serie_lkh}
\end{equation}
The expression can be integrated to give
\begin{equation}
\log P= \sum_K \left[ -\frac{1}{2}\log a_K - \frac{SSE_K}{2} + \frac{1}{2} \log \pi+\log C_K \right].  \label{eqn:cost_series_flat_prior}
\end{equation}
%
\subsection{Generalization to multivariate normal distributions}
\label{app:datasegmentation:cost}
We assume to have a linear regression model that assesses the
relation between the dependent variables \(y_i\) (the data) and the
\(N\)-vector of regressors \(x_i\) (unknowns).
This relation is written in vector form as
\begin{equation}
    y = H x + \varepsilon.   \label{eqn:linear_regression}
\end{equation}
Thus, \(y\) is an \(M\)-vector, \(H\) an \(M \times N\) matrix, \(\varepsilon\) an
\(M\)-vector of errors, and \(x\) an \(N\)-vector.  The errors are assumed to be
independently normally distributed with mean 0 and a variance \(\sigma\),
that is
\begin{equation*}
\varepsilon \sim N(0_M,[\sigma^2]),
\end{equation*}
where \(0_M\) is an \(M\)-vector of zeroes and \([\sigma^2]\) is an \(M\times M\)
diagonal matrix whose diagonal is \(\{\sigma_1^2,\sigma_2^2,\ldots,\sigma_M^2 \}\).
Using the properties of the multivariate normal distribution, it
follows that the likelihood function is given by
\begin{equation}
p(y| x, \sigma)= \frac {1} {2 \pi^{M/2}  \prod_{i=1}^{M} {\sigma_i} } e^{-\frac{1}{2} (y-Hx)'[\sigma^{-2}] (y- Hx)}.
\end{equation}
The likelihood function can be rewritten in terms of ordinary least-squares quantities:
\begin{equation*}
\begin{split}
& \hat {x}= (H'[\sigma^{-2}]H)^{-1} H^T[\sigma^{-2}]y  & \ \ \text{\small least-squares solution} \\
& SSE=(y-H \hat {x})^T[\sigma^{-2}] (y-H \hat {x}) & \ \ \text{\small sum of squared errors}.
\end{split}
\end{equation*}
If we take the highest likelihood value, 
\begin{equation}
\log P= \log C - \frac{SSE}{2} \mathrm{~with~} C=\frac {1} {2 \pi^{M/2}  \prod_{i=1}^{M} {\sigma_i} }. \label{eqn:cost_data_marginalized}
\end{equation}
If the variable \(x\) is marginalized by choosing the flat un-normalized prior,
it yields for the marginal posterior after integration
\begin{equation}
\log P=\log C + \frac{1}{2} \log \frac{(2 \pi)^p}{det (H'[\sigma^{-2}]H )} -\frac{SSE}{2}.  \label{eqn:cost_data_flat_prior}
\end{equation}
For \(M=\Nseries\),~\(N=1\) and \(H\) a matrix with elements \(H_{ij}=1\), we recover Eq.~\ref{eqn:max_serie_lkh}
and Eq.~\ref{eqn:marg_serie_lkh}, with 
\begin{equation*}
a=   det (H'[\sigma^{-2}]H )= \sum_{i=1}^{N} {\frac{1}{\sigma_i^2}   }. 
\end{equation*}
%
\section{Common model selection criteria}
\label{app:penalty}
Other penalty cost functions are possible, the most widely known criteria  for model
selection among a class of parametric models with different numbers of
parameters are \(AIC\)  \citep{Akaike74}  and \(BIC\) \citep{Schwartz78}. 
Given \(\Likeh\), the likelihood of the data, their expressions are 
\(AIC= 2k - 2\log(\Likeh)\), \(BIC = {-2 \cdot \log{\Likeh} + k \log(M) }\).
Here  \(M\) is the number of data and \(k\) the number of parameters used in the model.
Both \(AIC\) and \(BIC\)  fit the general form given in Eq.~\ref{eqn:objf} with
 \((\beta=2)\) and \((\beta=\log{M})\).
The penalty term of \(BIC\) is more stringent than the penalty term of \(AIC\)
(for \(M > 8\), \(k \times \log M\) exceeds \(2k\)). The \(AIC\) formula is based on an asymptotic
behavior with \(M\) large and a small number of parameters \(k\) and \(BIC\) on an
asymptotic approximation to a transformation of the Bayesian posterior
probability of a candidate model. Thus, both are issued from asymptotic
approximation and may have some drawbacks in some range of parameters.
Modified versions of
these criteria have also been developed to correct their main
weakness. 
The Hannan-Quinn information criterion (\(HQC\)) \citep{HannanQuinn79} proposes something
intermediate with (\(HQC= {-2 \log{\Likeh} + 2k \log(\log M) }\)) and (\(\beta=2\log{(\log{M})}\)).
%
\section{Mathematical utilities}
\label{app:lemna}
The formulas below avoid recomputing the inverse of matrix from scratch when a small rank adjustment is performed on the matrix.
%
\subsection{Sherman-Morrison-Woodbury formula - inversion lemma}
\label{app:lemna:woodbury}
Suppose that one aims to invert the matrix \(A + UCV\). Here \(A\),
\(U\), \(C\), and \(V\) are \(n\times n\), \(n\times k\), \(k\times k\)
and \(k\times n\) matrices, respectively.
\(UCV\)  is called rank-\(k\) update (or correction) of \(A\).
If \(C\) has a much smaller dimension than \(A\), it is more efficient to use the
Sherman-Morrison-Woodbury formula than to invert \(A+UCV\) directly.
Now suppose that \(A^{-1}\) has already been computed and that \(C\) and
\(C^{-1} + VA^{-1}U\) are invertible.
The Sherman-Morrison-Woodbury formula says that the inverse of a
rank-\(k\) correction of some matrix can be computed by applying a rank-\(k\)
correction to the inverse of the original matrix.  Alternative names
for this formula are the matrix inversion lemma or simply the Woodbury
formula. The Woodbury matrix identity is
\begin{equation*}
    \left(A+UCV \right)^{-1} = A^{-1} - A^{-1}U \left(C^{-1}+VA^{-1}U \right)^{-1} VA^{-1}.
\end{equation*}
The determinant of \(A + UCV\) is obtained through the relation:
\begin{equation*}
    \det(A+UCV) = \det(C^{-1} + V A^{-1} U)\det(C)\det(A).
\end{equation*}
%
\subsection{Inverse of a partitioned matrix}
\label{app:lemna:partition}
Let the \(N\times N\) matrix \(A\) be partitioned into a block form as below, where
\(T\) and \(W\) are square matrices of size \(t \times t\) and \(w \times w\),
respectively (\(t+w=N\)).  Matrices \(U\) and \(V\) are not necessarily square,
and have size \(t \times w\) and \(w \times t\), respectively.  Let
matrices \(T\), \(W\) and \(Q=W-V T^{-1}\) being invertible. The inverse of \(A\)
can be written as
\begin{equation*}
A^{-1}=
\begin{bmatrix}  T &  U \\  V & W \end{bmatrix}^{-1} 
= \begin{bmatrix} T^{-1} + T^{-1} U Q^{-1} V T^{-1} & -T^{-1} U  Q^{-1} \\ -Q^{-1} V T^{-1} & Q^{-1} \end{bmatrix} . 
\end{equation*}
The determinant of \(A\) is obtained through the relation:
\begin{equation*}
    \det(A)=det(T)\det(W-V T^{-1} U) = det(W)\det(T-U W^{-1} V).
\end{equation*}
%
\subsection{Application}
\label{app:lemna:application}
We have to solve system \(Hx=y\) in the least-squares sense; \(y\) is the right-hand side, \(H\) a matrix and \(x\) 
the solution, see also Appendix \ref{app:phystomath}. Now,
we need to update the solution \(x\) of the system of equations
\(Ax=(H^TH)x=H^T y\) after adding or deleting some
columns to \(H\). Here \(H\) is a matrix of size \(M \times N\), \(x\) and \(y\) are
vectors of length \(N\) and \(M\), respectively.  We first derive the
expression of the updated inverse of \(A\). We assume that the inverse of \(A\) is already computed
and denote by \(A_*\) the updated corresponding
matrix. To simplify, we just plug the formula for 1 column added or
suppressed.  First we add \(N_{add}\) columns to \(H\) such that \(H_*=[H v]\),
\(v\) is a matrix of size \(M \times 1\) representing the newly added
column, then
\begin{equation*}
A_*=
\begin{bmatrix}  H^T \\  v^T \end{bmatrix} \begin{bmatrix}  H &  v \end{bmatrix} = \begin{bmatrix}  H^T H & H^T v \\  v^T H & v^T v \end{bmatrix}.
\end{equation*}
The inverse of this partitioned matrix is
(App.~\ref{app:lemna:partition})
\begin{equation*}
\begin{split}
A_*^{-1}&=\begin{bmatrix}  H^T H & H^T v \\  v^T H & v^T v \end{bmatrix}^{-1}\\
        &=\begin{bmatrix} A^{-1}+A^{-1}H^T v \Delta^{-1} v^T H A^{-1} & A^{-1} H^T v \Delta^{-1} \\ -\Delta^{-1} v^T H A^{-1}  & \Delta^{-1} \end{bmatrix}\\
        &=\begin{bmatrix} F & -u_3 \\  -u_3^t  & \Delta \end{bmatrix},
\end{split}
\end{equation*}
where \(\Delta=v^T v-v^T H A^{-1} H^T v\). 
Algorithm~\ref{alg:addcol} performs the operation without accessing \(A^{-1}\) explicitly. We use the MATLAB notation ``\(\backslash\)'' to denote the solution of a linear system (\(``x=A\backslash b" \equiv ``x=A^{-1}b"\)): it emphasizes the fact that the system is solved without forming the inverse of the matrix of the system.
\begin{algorithm}[ht]
\caption{Implicit rank-1 update of \ensuremath{A=(H^T H)^{-1}} after adding a column \ensuremath{v} to \ensuremath{H} at position \ensuremath{j}.}\label{alg:addcol}
\begin{algorithmic}[1]
\STATE \(u_1 = H^T v\)
\STATE \(u_2 = A\backslash u_1\)
\STATE \(\Delta = [v^T v -u_1^T u_2]^{-1}\) 
\STATE \(u_3    = u_2 \Delta \)
\STATE \(A_*^{-1} \leftarrow \begin{bmatrix} A^{-1}+\Delta u_2^T u_2  & -u_3 \\  -u_3^t  & \Delta \end{bmatrix}\) \COMMENT{Not explicitly formed}
\STATE \(z(1:N+1)=[H v]^Ty\) \COMMENT{A vector of length \(n+1\)}  
\STATE \(  x_* \leftarrow \begin{cases}x(1:N)=x+u_3\left(u_2^T z(1:N)-z(N+1)\right) \\
                 x(N+1)=-u_3^T z(1:N)+\Delta y(N+1) \end{cases}\) \COMMENT{Update x without using \(A_*^-1\) explicitly} 
\STATE \(y_*^{predicted} \leftarrow  [H v] x_*\) \COMMENT{Predicted data}
\STATE Permute column \(J\) and row \(J\) of \(A_*^{-1}\) to last column and last row \\
\STATE Permute element \(J\) of  \(x_*\) to last element
\end{algorithmic}
\textbf{Output:} Updated least-squares solution and predicted data \(x_*, y_*^{predicted}\)
\end{algorithm}
If we need to remove a column from \(H\), the algorithm is described in algorithm~\ref{alg:subcol}.
\begin{algorithm}[ht]
\caption{Implicit rank-1 update of \ensuremath{A=(H^TH)^{-1}} after suppressing column \ensuremath{j} of \ensuremath{H}.}\label{alg:subcol}
\begin{algorithmic}[1]
\STATE \(e_j=0\)  and \(e_j(j)=-1\) \COMMENT{\(e_j\) is a unit vector}
\STATE \(A u_3=e_j\) \COMMENT{One system to solve to have \(u_3=A^{-1} e_j\)}
\STATE \(\Delta=(e_j^T u_3)^{-1}\) \COMMENT{Obtain element \(A^{-1}_{j,j}\)} 
\STATE \(A_*^{-1} \leftarrow  A^{-1}(1:j-1,j+1:N, 1:j-1,j+1:N)-u_3 \Delta u_3^T\) \COMMENT{Not explicitly formed}
\STATE \(z=H^Ty\)
\STATE \(w=x-u_3 e_j^T z\)
\STATE \(x_* \leftarrow w-u_3 \Delta z\)
\STATE \(y_*^{predicted}  \leftarrow H x_*\)
\end{algorithmic}
\textbf{Output:} Updated least-squares solution and predicted data \(x_*, y_*^{predicted}\) 
\end{algorithm}
%
\subsection{Adding \(N_{add}\) columns and suppressing \(N_{sub}\) columns}
\label{app:lemna:application2}
The two algorithms described above are also applicable without loss of
generality when several columns are added or
removed. Algorithm~\ref{alg:update} first adds \(N_{add}\) columns and
then suppresses \(N_{sub}\) columns from the transfer matrix \(H\) and updates
the solution \(x_*\) and predicted data \(y_*^{predicted}\).
\onecolumn
\begin{algorithm}[ht]
\begin{tiny}
\caption{\ensuremath{A=(H^TH)^{-1}} update after adding \ensuremath{N_{add}} and suppressing \ensuremath{N_{sub}} columns from \ensuremath{H}.} \label{alg:update}
\begin{tabular}[h]{lp{0.8\textwidth}}
{\bf Input:}  & \(m,n\) size of the system \\
&         \(H\): Transfer matrix (\(m\times n\)) to be updated \\
&         \(y(m)\):  data \\
&         \(jsubmin, jsubmax\): contiguous column of \(H\) to be suppressed (\(Nsub=jsubmax-jsubmin+1\)) \\
&         \(jaddmin, jaddmax\): contiguous column of \(H\) to be added (\(Nadd=jaddmax-jaddmin+1\)) \\
&         \(v(m,N_{add})\): columns to be added to \(H\) \\
&         \(x\): solution to \(Ax=(H^TH) x = H^Ty\) \\
&         \(z(1:n)=H^Ty  \& z(n+1:n+N_{add})=v^Ty\) \\
\end{tabular}
\algsetup{linenosize=\tiny}
\begin{algorithmic}[1]
\STATE \COMMENT{\textbf{Add \(N_{add}\) columns from \(jaddmin\) to \(jaddmax\) (append them first to position \(n+1:n+N_{add})\) }}
\STATE \(u_1=x^Tv\)
\STATE \(A u_2 = u_1\) \COMMENT{Solve \(u_2=A^{-1}u_1\)}
\STATE \(D_i=v^Tv-u_1^Tu_2\) \COMMENT{Symmetric matrix of order \(N_{add}\)}
\STATE \(D_a=D_i^{-1}\)
\STATE \(u_3=u_2 D_a\)
\STATE \(x_a(1:n)=x+u_3 (u_2^T z(1:n)-z(n+1:n+N_{add}))\) \COMMENT{Non-permuted solution after addition of \(N_{add}\) columns to \(H\)}
\STATE \(x_a(n+1:n+N_{add})=-u3^T z(1:n,1)+D_a z(n+1:n+N_{add},1)\) \COMMENT{(same)}
\STATE \COMMENT{\textbf{Suppress \(N_{sub}\) columns from \(jsubmin~to~jsubmax\)}}
\STATE \(e_j(n+N_{add},N_{sub})=0\)
\FOR{\(i=1\) \TO \(N_{sub}\)}
\STATE \(e_j(jsubmin+i-1,i)=1\)
\ENDFOR
\STATE \(A w =e_j\) \COMMENT{Solve \(w=A^{-1} e_j\)} 
\STATE \(u_{3s}(1:n,1:N_{sub})=w+u_3(u_2^T e_j(1:n,1:N_{sub})-e_j(n+1:n+N_{add},1:N_{sub}))\)
\STATE \(u_{3s}(n+1:n+N_{add},1:N_{sub})=-u_3^T e_j(1:n,1:N_{sub})+D_ae_j(n+1:n+N_{add},1:N_{sub})\)
\STATE \( D_i=e_j^T u_{3s}\) \COMMENT{\(D_{sub}\) is a symmetric matrix of order N\(_{sub}\)}
\STATE \( D_{sub}=D_i^{-1}\)
\STATE \(z=x_a-u_{3s} e_j^Tz\)
\STATE \(x_*=z-u_{3s} D_{sub} e_j^T z\) 
\STATE \(y_*=H x_*(1:n)+v x_*(n+1:n+N_{add})\)
\STATE \(x_*=x_*([1:jsubmin-1~ jsubmax+1:n+N_{add}])\)  \COMMENT{After suppressing elements jsubmin:jsubmax}  
\STATE \(x_*=x_*([1:jaddmin-1~ n-N_{sub}+1:n+N_{add}-N_{sub} ~jaddmin:n-N_{sub}])\) \COMMENT{Permute the output solution \(x_*\)} 
\end{algorithmic} 
\textbf{Output:} Updated least-squares solution and predicted data: \(x_*(n+N_{add}-N_{sub}), y_*(m)\) 
\end{tiny}
\end{algorithm}
%
\onecolumn
\section{`Pseudo-codes'}
The codes (simplified versions) for the 'image-space' and 'data-space' algorithms are written in IDL and Fortran,
the time series of GX 339-4 (\ref{sec:results:imagespace:timeseries}) and the data for V0332+53 (\ref{sec:results:dataspace:framework}) that are to be used as input data to these codes can be downloaded at http://sigma-2/integral/algorithms.
These codes and input data allow one to reproduce the main features of Figures ~\ref{fig:lc_gx339_image} and ~\ref{fig:datasegv0332}.
%
\subsection{`Image-space' algorithm}
\label{app:imagespace:algo}
\begin{algorithm}
\begin{tiny}
\caption{Data best partitioning} \label{algo:imagespace}
{\bf Input:} \\
\begin{tabular}[h]{lll}
{~} & L: & data length\\
    & x(1:L) :    & dataset of the form \((x_1,x_2,\ldots,x_L)\)\\
    & \(\beta\) : & penalty parameter\\
\end{tabular}
\algsetup{linenosize=\tiny}
\begin{algorithmic}[1] 
\STATE last(1:L)=0 ; best(1:L)=0.0 \COMMENT{Initialization}
\FOR[ Loop on the data subset 1 to R] {R=1 \TO L} 
     \STATE cost(1:n)=\(\mathcal{C}\)(x,R), see App. \ref{app:segmentation:cost} \COMMENT{Compute cost(i)=\(\mathcal{C}(x_{i:R})\) }
     \STATE F(1)=cost(1)+\(\beta\) \COMMENT{Segment starting at \(i=1\) and ending at \(R\) : \( F(0)+\mathcal{C}(x_{1:R}) \)}
     \FOR{i=2 \TO R}
          \STATE F(i)=best(i-1)+cost(i)+\(\beta\) \COMMENT{i.e. \( F(t)+\mathcal{C}(x_{t+1:R}) \) with \(t=i-1\)} 
     \ENDFOR
     \STATE [best(R),last(R)]=minloc(F(1:R))\COMMENT{minimum value and its location} 
\ENDFOR
\STATE \COMMENT{The vector  \(last\) gives the start point of individual segment of the optimal partition in the following way. Let \(\tau_{m}\)=\(last\)(L), \(\tau_{m-1}\)=\(last\)(\(\tau_{m}\)-1), etc. The last segments contains cells (data points)  \(\tau_{m},\tau_{m}+1,\ldots,L\), the next to last contains cells \(\tau_{m-1},\tau_{m-1}+1,\ldots,\tau_{m}-1\) and so on.}
\STATE  index=last(n) and cpt=[] \COMMENT{Change points storage}
\WHILE {index \(>\) 1}
   \STATE cpt=[index cpt]
   \STATE index=last(index-1)
\ENDWHILE
\end{algorithmic}
\textbf{Output:} \\
\begin{tabular}[h]{lll}
{~} & \(cpt=[1=\tau_0,\tau_1,\ldots,\tau_{m},\tau_{m+1}=L+1]\) : & change points\\
\end{tabular}
\end{tiny}
\end{algorithm}
%
\subsection{`Data-space' algorithm}
\label{app:dataspace:algo}
\begin{algorithm}[ht]
\begin{tiny}
\caption{`Data-space' algorithm} \label{algo:dataspace}
{\bf Input:} \\
\begin{tabular}[h]{lll}
{~}  & \(nofp\) :  & number of exposures \\
     & \(nofs\) :  & number of sources \\ 
     & \(ndata\) :  & number of data points \\
     & \(Hzero(ndata:nofs)\) : & weighted initial transfer matrix (\( H^0 \equiv \frac {H^0_{ij}} {\sigma_i} \)) \\
     & \(y(ndata) \) : & weighted data ( \(  y_i \equiv \frac {y_i} {\sigma_i} \)) \\
     & \(index_-data(nofp+1)\) : & vector of indexes such that  \( y(index_-data(p):index_-data(p+1)-1)) \) are data points of exposure p \\
     & \(beta\) : & penalty parameter beta \\
     & \(maxiter\) : & number of maximum subiterations \\
\end{tabular}
\algsetup{linenosize=\tiny}
\begin{algorithmic}[1] 
\STATE  \(lastarr(1,1:nofs)=1\)  \COMMENT{Array of change points}
\STATE  \(best(1:nofp)=1e77\)  \COMMENT{Initialize the cost function for each iteration}
\FOR[Add a new set of data at each iteration R]{R=2 \TO nofp}
\STATE \(converge=\TRUE\) \COMMENT{The cost function at iteration is improved}
\STATE \(iter=0\) \COMMENT{Subiteration counter}
\STATE \(m=index_-data(R+1)-1\) \COMMENT{Total number of data points used at iteration R}
\STATE \(lastarr(R,1:nofs)=lastarr(R-1,1:nofs)\) \COMMENT{Extend the previous segments}
\WHILE {\(converge\)}
\STATE \COMMENT{Compute the cost at iteration R}
\STATE \COMMENT{\(next_-bestJ(1:nofs)\): best cost function for each source}
\STATE \COMMENT{\(next_-lastrJ(1:nofs)\): and the corresponding change-point locations}
\STATE \COMMENT{\(cost_-without_-new_-chgpt\): cost function with no new change points added}
\STATE \COMMENT{\(nseg_-array(1:nofs)\): the number of segments per source with no change points added}
\STATE \(N=sum(nseg_-array(1:nofs))\) \COMMENT{Total number of segments/column of matrix H}
\IF [If the system of equations is not underdetermined]{\(N <= m\)}  
\STATE \(iter=iter+1\)
\STATE \([bestRj,isonum]=minloc(next_-bestJ)\)  \COMMENT{Minimum cost value, its location in terms of source number}
\IF [Addition of a new change point improves the cost function]{\(bestRj < cost_-without_-new_-chgpt\)}  
\STATE \(best(R)=bestRj\)
\STATE \(lastarr(R,isonum)=lastRj(isonum)\) \COMMENT{Update the array of change points}
\ELSE
\STATE \(converge=\FALSE\) \COMMENT{No improvement of the cost function}
\STATE \(best(R)=cost_-without_-new_-chgpt\)
\ENDIF
\IF [Maximum subiterations reached]{\(iter > maxiter\)}  
\STATE\(converge=\FALSE\) \COMMENT{Next iteration}
\ENDIF
\ELSE
\STATE \(converge=\FALSE\) \COMMENT{Next iteration}
\ENDIF
\ENDWHILE
\ENDFOR
\end{algorithmic}
\textbf{Output:} \\
\begin{tabular}[h]{ll}
\(lastarr(1:nofp,1:nofs)\) : & vector \(lastarr(nofp,Jnum)\) is backtracked/peeled off to obtain the location of the change points of source \(Jnum\) \\
\end{tabular}
\end{tiny}
\end{algorithm}

\newpage

\begin{algorithm}[ht]
\begin{tiny}
\caption{'Data space' cost function computation} \label{algo:datacost}
{\bf Input:} \\
\begin{tabular}{lll}
{~} & \(R\) :     &     iteration number/number of used exposures  \\
    & \(nofp\) :  &     number of exposures \\
    & \(nofs\) :  &     number of sources \\ 
    & \(ndata\) : &     number of data points \\
    & \(Hzero(ndata:nofs)\) : &  weighted initial transfer matrix (\( H^0 \equiv \frac {H^0_{ij}} {\sigma_i} \)) \\
    & \(y(ndata) \) : & weighted data ( \(  y_i \equiv \frac {y_i} {\sigma_i} \)) \\
    & \(index_-data(nofp+1)\): &  vector of indexes such that  \\
    &                              & \( y(index_-data(p):index_-data(p+1)-1)) \) \\
    &                              & are data points of exposure p \\
    & \(beta\)    : &  penalty parameter beta \\
    & \(maxiter\) : & number of maximum subiterations \\
    & \(lastarr(1:nofp,1:nofs)\) : & vector \(lastarr(nofp,Jnum)\) is backtracked/peel-off to obtain the location of the change points of source \(Jnum\) \\
\end{tabular}
\algsetup{linenosize=\tiny}
\begin{algorithmic}[1] 
\FOR [Find the \ensuremath{nseg} segments in the array \ensuremath{lastarr(1:R,iso)}] {\(iso=1\ \TO\ nofs\)}
\STATE \(nseg_-array(iso)=nseg\)
\ENDFOR
\STATE \(m=index_-data(R+1)-1\)  \COMMENT{Total number of data points used at iteration \(R\)}
\STATE\(valid_-source(1:nofs)=1\)  \COMMENT{All sources are to be ``partitioned''} 
\STATE \COMMENT{Optionally, by setting \(valid_-source(J)=0\) and  \(nseg_-array(J)=0\), source \(J\) is not 
used to build the matrix \(H\); the source makes the system of equations not well-conditioned or the source is 
detected with a too low efficiency (low surface area ( \(cm^2\)) projected on the camera).}
\STATE \(N=sum(nseg_-array)\) \COMMENT{Number of columns in matrix \(H\)}
\IF [Return if this system is underdetermined] {\(N >  m\)}
   \STATE \(cost_-without_-new_-chgpt=0.\)
   \STATE \(next_-bestJ=1.0d+77\)
   \STATE \(next_-lastrJ=-1 \)
   \STATE \(return \)
\ENDIF
\STATE \(H(1:m,1:N)=0.\)  \COMMENT{Initialization of the matrix \(H\)}
\STATE \(kw=0\) 
\FOR [Build the transfer function] {\(iso=1\ \TO\ nofs\)}
   \STATE set \(ka(1:nseg+1)\) \COMMENT{Peel-off the vector \(last(1:R,iso)\), the segment j starts at ka(j) and ends at ka(j+1)-1}
   \IF [Consider only the list of valid  sources] {\(valid_-source(iso) == 1\)}
      \FOR [Expand column \ensuremath{H(1:m,iso)}]{\(j=1\ \TO\ nseg\)}
         \STATE \(kw=kw+1\)
         \FOR {\(i=ka(j)\ \TO\ ka(j+1)-1\)}
            \STATE \(ilow=index_-data(i)\); \(iupp=index_-data(i+1)-1\) 
            \STATE \(H(ilow:iupp,kw)=Hzero(ilow:iupp,iso)\)
         \ENDFOR
      \ENDFOR
   \ENDIF         
\ENDFOR
\STATE \COMMENT{ Solve the system of equations and return the auxiliary quantities needed to compute the cost function}
\STATE \(cost_-without_-new_-chgpt=-log(P)\)  \COMMENT{Cost function with \(N\) segments}
\STATE \(irowmin(1:nofs)=1\) \COMMENT{Contains  the starting of the ``loop i'' for each sources}
\STATE \COMMENT{Optionally modify the starting values \(irowmin(1:nofs)\) of the ``loop i''}
\FOR [Identify the best new change point for each source]{\(Jnum=1\ \TO \ nofs\)}
    \STATE \COMMENT{Loop on source (Algorithm~\ref{algo:loop_on_source}) returns \(next_-bestJ(1:nofs)\) and \(next_-lastrJ(1:nofs)\)}
\ENDFOR [Loop on source \(Jnum\)]
\end{algorithmic}
{\bf Output:}\\
\begin{tabular}[h]{ll}
 \(next_-bestJ(1:nofs)\):     & best cost function for each sources \\
 \(next_-lastrJ(1:nofs)\):    & and the corresponding vector of change points \\
 \(cost_-without_-new_-chgpt\): & cost function with no new change points added \\
 \(nseg_-array(1:nofs)\):       & the number of segments per source with \\
                              &  no new change points added\\
\end{tabular}
\end{tiny}
\end{algorithm}
%
\begin{algorithm}[ht]
\begin{tiny}
\caption{``Loop on sources''} \label{algo:loop_on_source}
{\bf Output:}\\
\begin{tabular}{lll}
{~} & \(next_-bestJ(1:nofs)\):  & best cost function for each source \\
    & \(next_-lastrJ(1:nofs)\): & and the corresponding change-point locations \\
\end{tabular}
\algsetup{linenosize=\tiny}
\begin{algorithmic}[1] 
\STATE \(next_-lastrJ(Jnum)=lastarr(R,Jnum)\)     \COMMENT{Best change-point location}
\STATE \(next_-bestJ(Jnum)1.0e77 \)               \COMMENT{and the default cost function}

\IF [If the source is to be ``partitioned''] {\(valid_-source(Jnum) = 1\)}
\STATE \(cost(1:R)=1.0e77 \)           \COMMENT{Initialize to high value}
%
\FOR [Test the change point located at \ensuremath{irow}] {\(irow=irowmin(Jnum)\ \TO \ R\)}
   \IF [] {\(irow > 1\)}
      \STATE \COMMENT{Identify the segments delimiter \(ka(1:nseg+1)\) and the number of segments \(nseg\) in array \(last(1:irow-1,Jnum)\)}
      \STATE \(ka(nseg+2)=R+1\) ; \(nseg=nseg+1\)  
      \STATE \(ka(1)=1\), \(ka(2)=R+1\) ; \(nseg=1\)
   \ELSE   
      \STATE \(ka(1)=1\), \(ka(2)=R+1\) ; \(nseg=1\) \COMMENT{Special case of a single segment \(irow = 1\)}
   \ENDIF
   \STATE \(N_-star=N-nseg_-array(Jnum)+nseg\)   \COMMENT{Number of segments of the new matrix \(H_-star\)}
   \IF [If the resulting system is not underdetermined] {\(N_-star \le m\)}
      \STATE \(Hj_-star(1:m,1:nseg)=0.0\) \COMMENT{Initialize the matrix \(Hj_-star\)}
      \STATE \(kw=0\)
      \FOR [Construct the new submatrix for source \ensuremath{Jnum}]{\(j=1\ \TO \ nseg\)}
          \STATE \(kw=kw+1\)
          \FOR {\(i=ka(j)\ \TO\ ka(j+1)-1\)}
             \STATE \(ilow=index_-data(i)(i)\) and \(iupp=,index_-data(i+1)-1\) 
             \STATE \(Hj_-star((ilow:iupp,kw)=Hzero(ilow:iupp,Jnum)\)
           \ENDFOR
      \ENDFOR
      \STATE set \(correct_-conf=1\) \COMMENT{This submatrix has some desired mathematical properties}
      \STATE \COMMENT{If this submatrix does not have some desired mathematical properties, 
                      set \(correct_-conf=0\), for example for a null column}
      \IF [Insert this submatrix in the place of the old submatrix to form the new matrix of the partition] {\(correct_-conf = 1\)}
          \STATE \(kmin=sum(nseg_-array(1:Jnum-1))\) ; \(kmax=kmin+nseg_-array(Jnum)\)
          \STATE \(H_-star(1:m,1:N_-star)=0\) \COMMENT{Initialize the matrix \(H_-star\)}
          \STATE \(H_-star(1:m,1:kmin)=H(1:m,1:kmin)\) \COMMENT{Original matrix \(H\)}
          \STATE \(H_-star(1:m,kmin+1:kmin+nseg)=Hj_-star(1:m,1:nseg)\) \COMMENT{Submatrix corresponding to source \(Jnum\)}
          \STATE \(H_-star(1:m,kmin+nseg+1:N_-star)=H(1:m,kmax+1:N)\) \COMMENT{Original matrix \(H\)}
          \STATE \COMMENT{ Solve the system equation and compute the auxiliary quantities to compute the cost function}
          \IF [If the system of equation is well conditioned or fulfills some constraints] {\(well_-conditioned_-system=1\)}
              \STATE \(cost(irow)=-log(P)\)  \COMMENT{Cost function with the \(N_-star\) segments}
          \ENDIF [End of\(well_-conditioned_-system=1\) ]

      \ENDIF     [End \(Correct_-conf=1\)]
    \ELSE   
      \STATE \COMMENT{Display optionally a WARNING because of a subdetermined system of equations: \(N_-star > m\)}
    \ENDIF [Loop \(N_-star < m\)]
\ENDFOR
\STATE \COMMENT{Locate the most probable change point in the array \(cost(irowmin(Jnum):R) \)}
\STATE \([next_-bestJ(Jnum),next_-lastrJ(Jnum)]=minloc(cost(irowmin(Jnum):R)\)) \COMMENT{Best cost value  and corresponding location}
\ENDIF [If the source is to be `partitioned' \(valid_-source(Jnum) = 1\)]
\end{algorithmic}
{\bf Output:}\\
\begin{tabular}[h]{ll}
 \(next_-bestJ(1:nofs)\) :  & best cost function for each sources \\
 \(next_-lastrJ(1:nofs)\) : & and the change-points position \\
\end{tabular}
\end{tiny}
\end{algorithm}
\end{appendix}

\end{document}